\documentclass[aps,prd,11pt,superscriptaddress,floatfix,nofootinbib,notitlepage,reprint,longbibliography
]{revtex4-1}
\usepackage{colortbl}
\usepackage[T1]{fontenc} 
\usepackage{lmodern}
\usepackage{hyperref}
\usepackage{graphicx}
\usepackage{bm}
\usepackage{amsmath}
\usepackage{amssymb}
\usepackage{dcolumn}
\usepackage{multirow}
\usepackage{color}
\usepackage[dvipsnames]{xcolor}
\usepackage{aas_macros}
\usepackage{fancyhdr}
\usepackage[papersize={8.5in,11in},top=2cm, bottom=1.5cm, left=1.5cm, right=1.5cm]{geometry}

\newcommand{\WW}[1]{#1}

\newcommand{\IRS}[1]{#1}

\newcommand{\msun}{\ensuremath{\mathrm{M}_\odot}}
\begin{document}
\title{Neutrinos from type Ia Supernovae: The gravitationally confined detonation scenario}

\author{Warren P. \surname{Wright}}
\email{wpwright@ncsu.edu}
\affiliation{Department of Physics,  North Carolina State University, Raleigh, North Carolina 27695, USA}

\author{James P. \surname{Kneller}}
\email{jpknelle@ncsu.edu}
\affiliation{Department of Physics,  North Carolina State University, Raleigh, North Carolina 27695, USA}

\author{Sebastian T. \surname{Ohlmann}}
\email{sebastian.ohlmann@h-its.org}
\affiliation{Heidelberger Institut f\"{u}r Theoretische Studien, Schloss-Wolfsbrunnenweg 35, D-69118 Heidelberg, Germany}

\author{Friedrich K. \surname{R{\"o}pke}}
\email{friedrich.roepke@h-its.org}
\affiliation{Heidelberger Institut f\"{u}r Theoretische Studien, Schloss-Wolfsbrunnenweg 35, D-69118 Heidelberg, Germany}
\affiliation{Zentrum f{\"u}r Astronomie der Universit{\"a}t Heidelberg, Institut f{\"u}r Theoretische Astrophysik, Philosophenweg 12, D-69120 Heidelberg, Germany}

\author{Kate \surname{Scholberg}}
\email{schol@phy.duke.edu}
\affiliation{Department of Physics, Duke University, Durham North Carolina  27708, USA}

\author{Ivo R. \surname{Seitenzahl}}
\email{ivo.seitenzahl@anu.edu.au}
\affiliation{Research School of Astronomy and Astrophysics, Australian National University, Canberra, Australia Capital Territory 2611, Australia} 
\affiliation{ARC Centre of Excellence for All-Sky Astrophysics (CAASTRO) 
}

\date{\today}

\begin{abstract}

Despite their use as cosmological distance indicators and their importance in the chemical evolution of galaxies, the unequivocal identification of the progenitor systems and explosion mechanism of normal type Ia supernovae (SNe~Ia) remains elusive. The leading hypothesis is that such a supernova is a thermonuclear explosion of a carbon-oxygen white dwarf, but the exact explosion mechanism is still a matter of debate. Observation of a galactic SN~Ia would be of immense value in answering the many open questions related to these events. One potentially useful source of information about the explosion mechanism and progenitor is the neutrino signal because the neutrinos from the different mechanisms possess distinct spectra as a function of time and energy. 
In this paper, we compute the expected neutrino signal from a gravitationally confined detonation (GCD) explosion scenario for a SN~Ia and show how the flux at Earth contains features in time and energy unique to this scenario. We then calculate the expected event rates in the Super-K, Hyper-K, JUNO, DUNE, and IceCube detectors and find both Hyper-K and IceCube will see a few events for a GCD supernova at 1 kpc or closer, while Super-K, JUNO, and DUNE will see events if the supernova is closer than ${\sim}0.3$ kpc. The distance and detector criteria needed to resolve the time and spectral features arising from the explosion mechanism, neutrino production, and neutrino oscillation processes are also discussed. 
The neutrino signal from the GCD is then compared with the signal from a deflagration-to-detonation transition (DDT) explosion model computed previously. We find the overall event rate is the most discriminating feature between the two scenarios followed by the event rate time structure. Using the event rate in the Hyper-K detector alone, the DDT can be distinguished from the GCD at 2$\sigma$ if the distance to the supernova is less than $2.3\;{\rm kpc}$ for a normal mass ordering and $3.6\;{\rm kpc}$ for an inverted ordering.
\end{abstract}

\maketitle

\section{Introduction \label{Introduction}}

The stellar explosions known as thermonuclear - or Type Ia - supernovae are important phenomena in astrophysics. SNe~Ia are the distance indicators \cite{Phillips1993,Phillips1999} which indicate the Universe is accelerating \cite{Riess1998,Schmidt1998,Perlmutter1999} and SNe~Ia are also major contributors to the chemical evolution of galaxies \cite{1997NuPhA.621..467N,2009ApJ...707.1466K} leading to notable changes in the rate at which, for example, oxygen and iron are enriched when the contribution from SNe~Ia becomes significant. But despite their importance, little is known about the progenitors of these supernovae. Only in a few cases for nearby SNe~Ia, such as SN 2011fe or SN 2014J, do pre-explosion images place strong constraints upon the luminosity of the progenitor system \cite{2011Natur.480..348L,2014ApJ...790....3K,2014MNRAS.442.3400N}. Nevertheless, various arguments point towards a paradigm that SNe Ia are the explosion of a carbon-oxygen white dwarf in a binary system that accretes sufficient mass from its companion to begin explosive carbon burning. The source of the accreted mass may be another, smaller, white dwarf (a double degenerate scenario) or a main sequence or giant star (a single degenerate scenario). The reader is referred to Maoz, Mannucci and Nelemans \cite{Maoz2014} and Ruiz-Lapuente \cite{Ruiz-Lapuente2014} for reviews and variations upon these scenarios. 

In addition to the debate over the identity of the companion, another debate is over the mechanism for the explosion itself. Popular mechanisms for the single degenerate scenario at the present time are the delayed detonation transition model (DDT) model \cite{DDToriginal}, the gravitationally confined detonation (GCD) model \cite{Plewa2004}, and the pulsational reverse detonation (PRD) model \cite{PRDoriginal}, though others also exist. We refer the reader to Hillebrandt \emph{et al.} \cite{Hillebrandt2013} for a recent review.
Determining the explosion mechanism from observation will be difficult. The unknown identity of the progenitor means  observatories will require close to full sky coverage, while the brevity of the explosion - simulations indicate the star becomes unbound within a few seconds - means sub-second integration times. Thus, it may not be until there is a galactic SN~Ia that the explosion mechanism can be determined observationally. The rate of SNe~Ia in the Galaxy is calculated to be $1.4^{+1.4}_{-0.8}$ per century by Adams \emph{et al.} \cite{2013ApJ...778..164A} and represents ${\sim}30\%$ of the total supernova rate.  Adams \emph{et al.} calculate the most probable distance to a Galactic SN Ia to be $d = 9\;{\rm kpc}$. 

At such close proximity, when the next SN Ia occurs within the Galaxy, not only can we exploit electromagnetic observations, it may also be possible to obtain information about the progenitor and explosion mechanism in the gravitational waves and the neutrino signal. The gravitational wave emission of the explosion (not the inspiral) has been considered by Falta \emph{et al.} \cite{2011PhRvL.106t1103F}
and Seitenzahl \emph{et al.} \cite{Seitenzahl2015a}. The neutrino emission was calculated by Kunugise and Iwamoto \cite{2007PASJ...59L..57K}, Odrzywolek and Plewa \cite{Odrzywolek2011a}, and Seitenzahl \emph{et al.} \cite{Seitenzahl2015a}. 
The study by Odrzywolek and Plewa \cite{Odrzywolek2011a} is particularly worth emphasizing because they showed how the neutrinos are capable of distinguishing between explosion mechanisms even when the electromagnetic output is identical. 

But as with many things neutrino related, detecting the neutrino signal from SNe Ia is not easy. In comparison with core-collapse supernovae, the flux of neutrinos at Earth from a SN Ia is approximately four orders of magnitude smaller for a source at the same distance and also the spectrum peaks in the range of $3\;{\rm MeV}$ \cite{2007PASJ...59L..57K} rather than the $10- 20\;{\rm MeV}$ for the neutrino spectrum from a core-collapse SN. The shift to lower energies reduces the number of events one expects in a detector because of the energy dependence of neutrino cross sections and, furthermore, detecting the events is harder because they occur much closer to, or below, detector thresholds. However one advantage of the neutrinos from SNe Ia compared to core-collapse SNe is that the neutrino signal is more reliably calculated. The densities in the core of SNe~Ia are not sufficient to trap the neutrinos so there is no neutrino transport to follow and there are also no neutrino self-interaction effects \cite{Duan:2006an,Duan:2006jv} because the neutrino density is not large enough. 

The advent of several next-generation neutrino detectors recently prompted Wright \emph{et al.} \cite{Wright2016} (hereafter called Paper I) to recompute the expected signal from a DDT SN Ia explosion 
including many effects which had not been taken into account previously. 
For their calculation, Wright \emph{et al.} took into account the full time and energy dependence of the emission and calculated the neutrino flavor transformation as a function of time and energy through the supernova and along several rays through the simulation in order to study the line-of-sight variability. They then processed the fluxes at Earth using the SNOwGLoBES detector event rate calculation software so as to compute the total and differential event rates in the Super-K, Hyper-K, JUNO, and DUNE detectors. A separate analysis was undertaken for the signal in the IceCube detector.
Their conclusion was that despite the difficulties, neutrino detectors are becoming so large and sensitive that the neutrinos from a DDT SN~Ia at the Galactic center can be detected. As the supernova is placed closer to Earth, features in the spectrum as a function of time and energy may be observed. 

The goal of this paper is to repeat the calculations of Paper I, but this time for the neutrino signal from the alternative scenario of a gravitationally confined detonation. 
The strategy is very similar to that found in Paper I and the outline of our paper is as follows. In Sec. \ref{sec:SneModel}, we provide an outline of the gravitationally confined detonation simulation we adopt. We then describe in Sec. \ref{sec:Production} the method by which the simulation is post-processed to compute the emitted neutrino spectra followed by Sec. \ref{sec:NeutrinoOscillation}, where we calculate the neutrino flavor transformation through the simulation as a function of time and energy along different rays. In Sec. \ref{sec:NeutrinoDetection}, we combine the emitted spectra and neutrino flavor transition probabilities to compute the flux at Earth and send these fluxes through the SNOwGLoBES software for representative examples of various neutrino detector technologies. The total and differential event rates we compute are investigated in order to determine how well the GCD supernova can be observed in neutrinos for a range of of distances to the event. 
Finally we compare the GCD and DDT neutrino signals in Sec. \ref{sec:Compare} in order to determine the distance/detector requirements necessary to distinguish the two scenarios. We conclude with Sec. \ref{sec:Conclusion}.

\section{Supernova Simulation \label{sec:SneModel}}

The SN explosion model used in our calculations is a gravitationally confined detonation (GCD) model. The explosion starts with off-center deflagration ignition in a near-Chandrasekhar-mass white dwarf (WD). This ignition is of a single bubble near the star's center and deflagration ash quickly `floats' to the surface creating a `plume'. The ash is prevented from escaping by the star's gravity. The gravitationally confined ash then proceeds to engulf the surface of the star converging upon a location opposite to the initial eruption point of the plume on the surface. The convergence of the ash leads to a compression which plausibly ignites a detonation \cite{2009ApJ...700..642S} that quickly propagates through the star unbinding it completely. This type of explosion mechanism was first discussed in \cite{Plewa2004} and the particular implementation of this scenario model used in our paper is taken from Seitenzahl \emph{et al.} \cite{Seitenzahl2016} where the full details can be found. In what follows, we provide a brief review. The explosion is modeled in three dimensions by use of the hydrodynamic SN code \textsc{Leafs} on a $512^3$ grid. This same code was also used to simulate a deflagration-to-detonation transition (DDT) model in \cite{Seitenzahl2012a} used in Paper I. The deflagration and detonation flames are modeled using a levelset approach (\citealp{Reinecke1999a}, for more details see also \citealp{Seitenzahl2016}). For reference, we summarize the basic properties of the two models in Table~\ref{table:explosion_models}.

\renewcommand{\arraystretch}{1.5}
\begin{table}[h]\centering
		\begin{tabular}{l  c  c  c  c}
			\cline{1-5}
			Model&IGE [\msun]&$^{56}$Ni [\msun]&IME [\msun] &E$_\mathbf{kin}$ [$10^{51}\,\mathrm{erg}$]\\ 
			\cline{1-5}
			GCD&$0.844 $&$0.742 $&$0.415$&$1.38$\\ 
			N100&$0.839 $&$0.604 $&$0.454$&$1.45$\\ 
			\cline{1-5}  
		\end{tabular}
   \caption{Compilation of the asymptotic kinetic energy (E$_\mathrm{kin}$) and nucleosynthesis yields for the Iron group elements (IGE), $^{56}\mathrm{Ni}$, and intermediate mass elements (IME) in solar masses (\msun) for the GCD \citep{Seitenzahl2016} and N100 \citep{Seitenzahl2012a} explosion models.}
	\label{table:explosion_models}
\end{table}
\renewcommand{\arraystretch}{1}

\renewcommand{\arraystretch}{0}
\setlength{\tabcolsep}{0pt}
\begin{figure*}
\begin{tabular}{cc}
\cellcolor{black}\includegraphics[height=0.13\textheight]{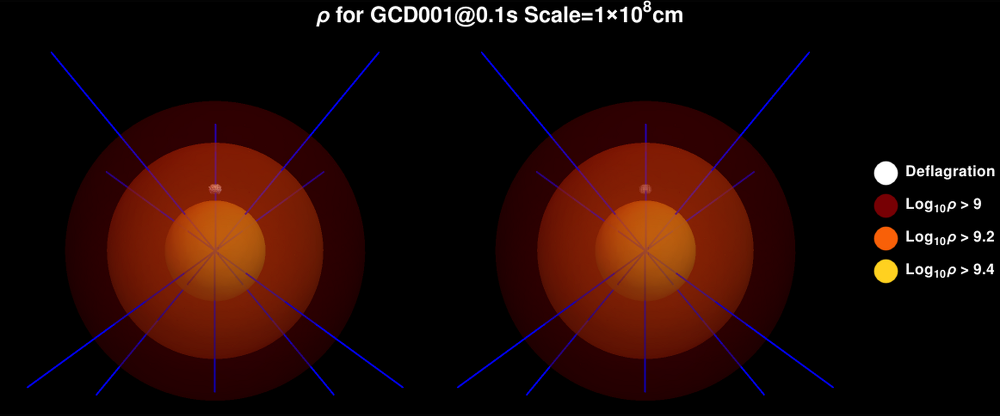} &
\cellcolor{black}\includegraphics[height=0.13\textheight]{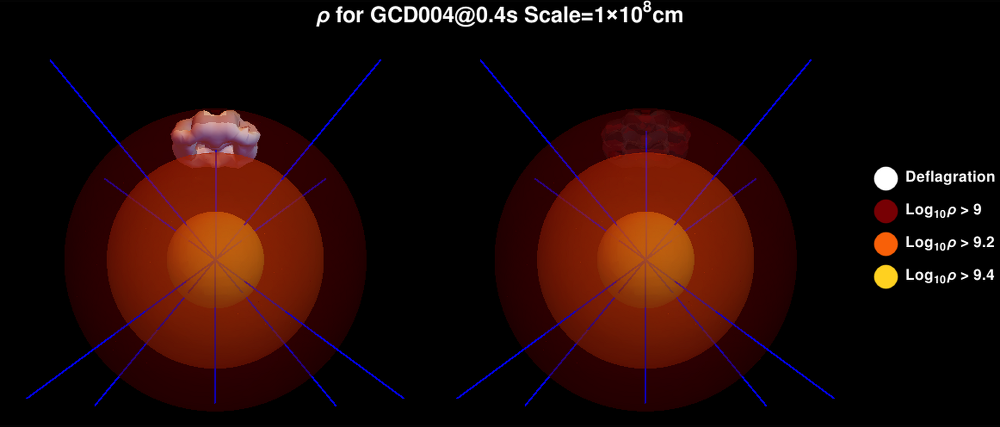} \\
\cellcolor{black}\includegraphics[height=0.13\textheight]{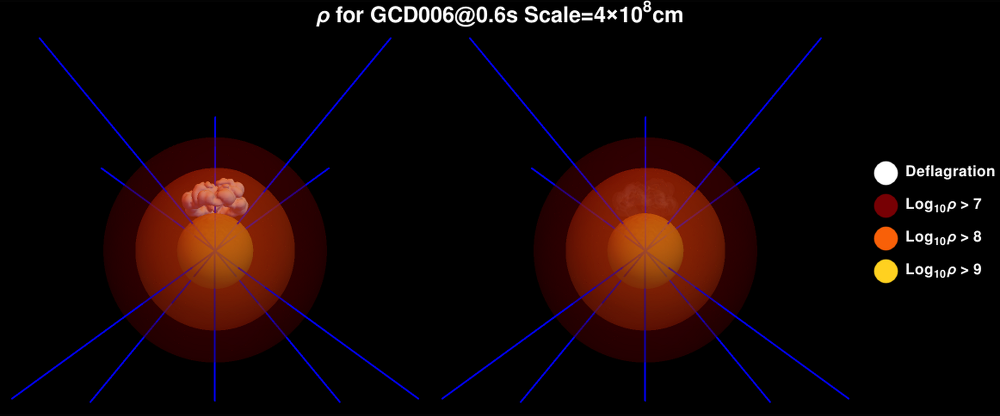} &
\cellcolor{black}\includegraphics[height=0.13\textheight]{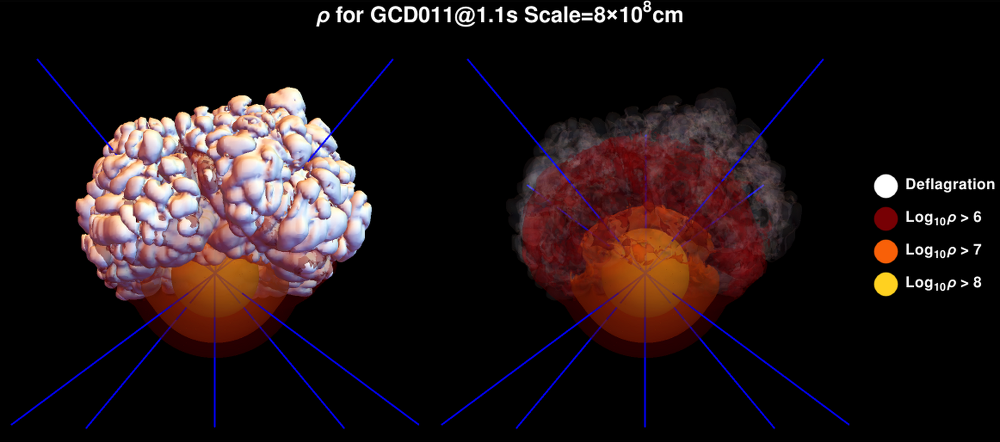} \\
\cellcolor{black}\includegraphics[height=0.13\textheight]{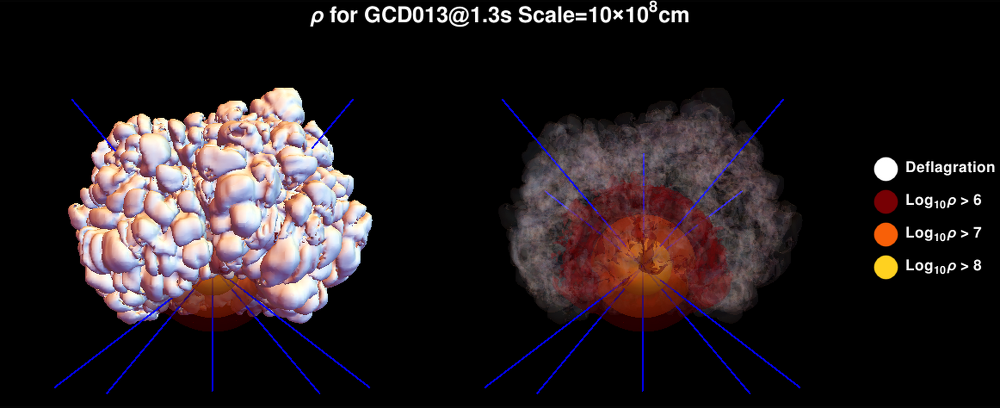} &
\cellcolor{black}\includegraphics[height=0.13\textheight]{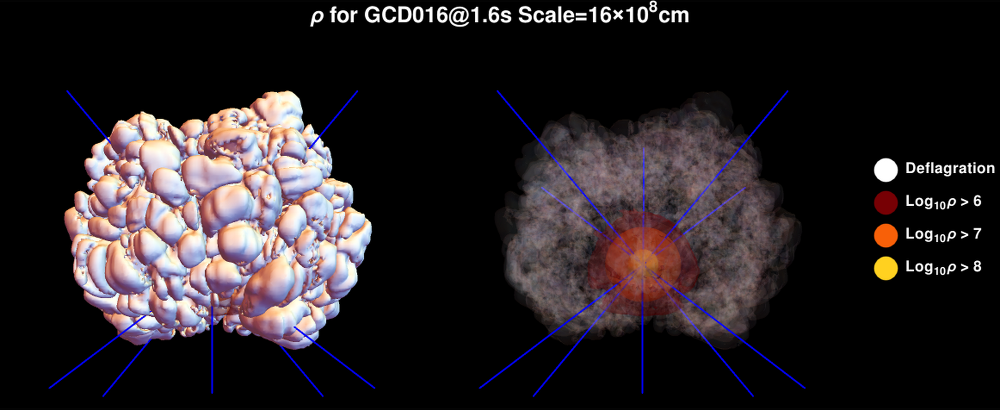} \\
\cellcolor{black}\includegraphics[height=0.13\textheight]{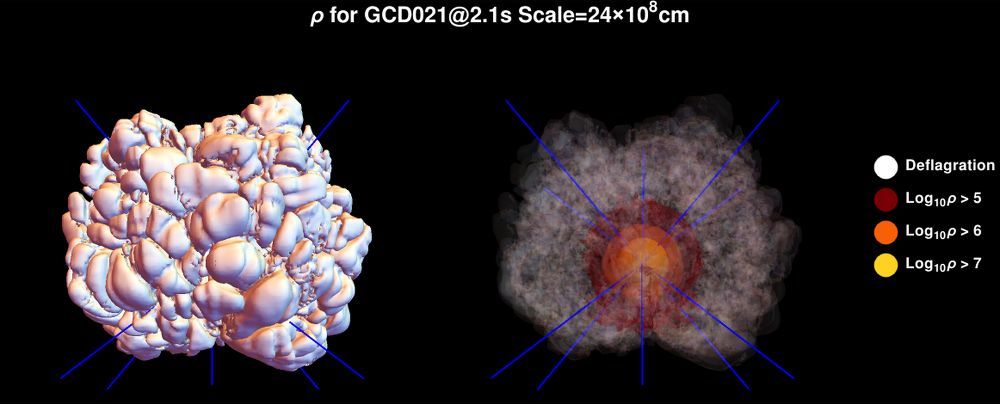} &
\cellcolor{black}\includegraphics[height=0.13\textheight]{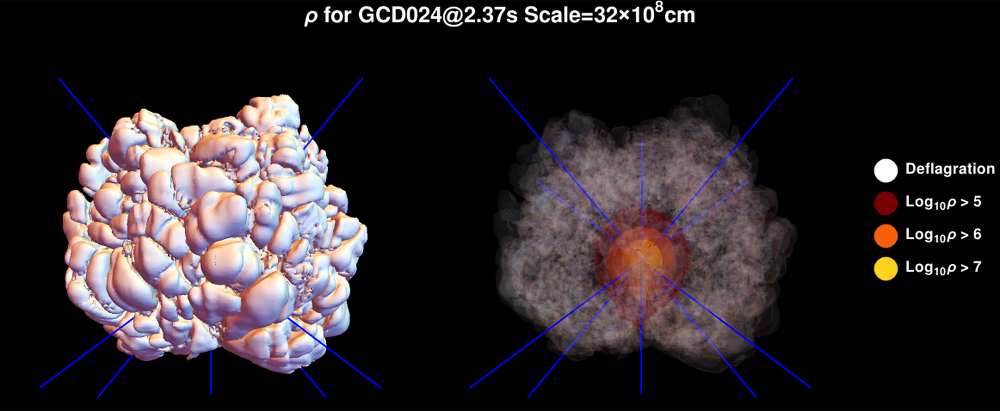} \\
\cellcolor{black}\includegraphics[height=0.13\textheight]{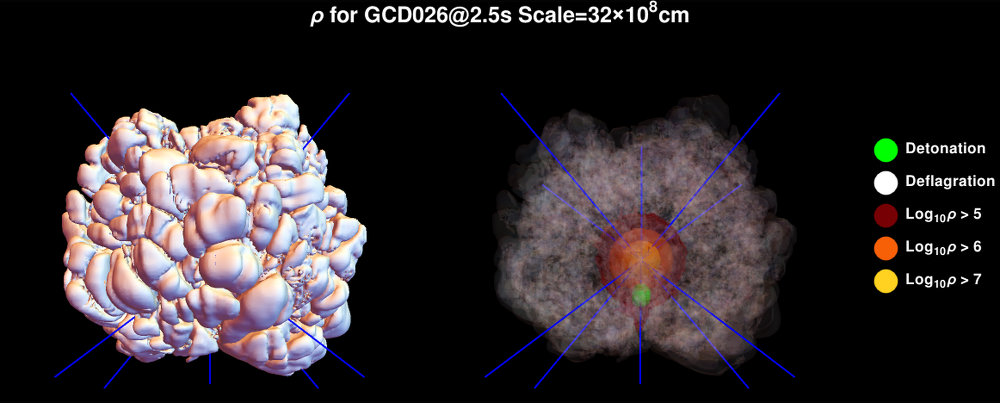} &
\cellcolor{black}\includegraphics[height=0.13\textheight]{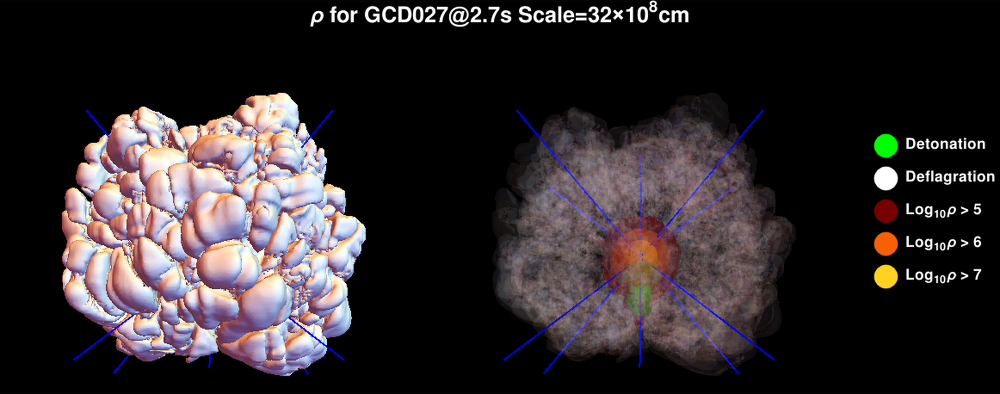} \\
\cellcolor{black}\includegraphics[height=0.13\textheight]{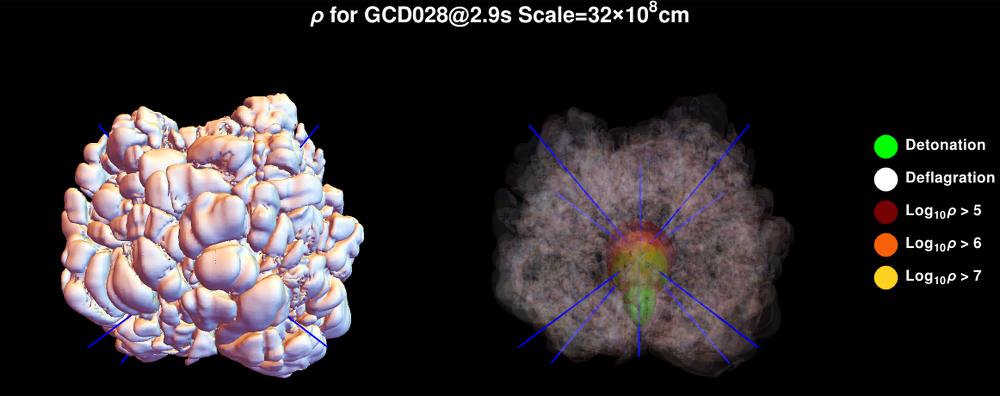} &
\cellcolor{black}\includegraphics[height=0.13\textheight]{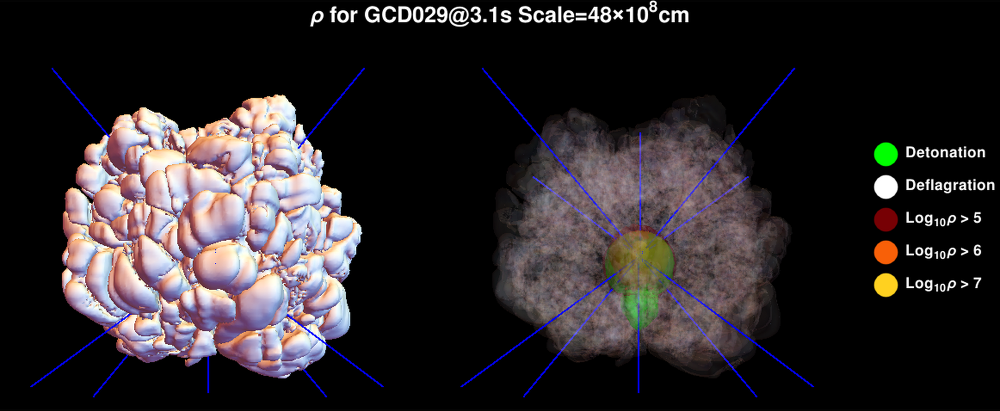} \\
\cellcolor{black}\includegraphics[height=0.13\textheight]{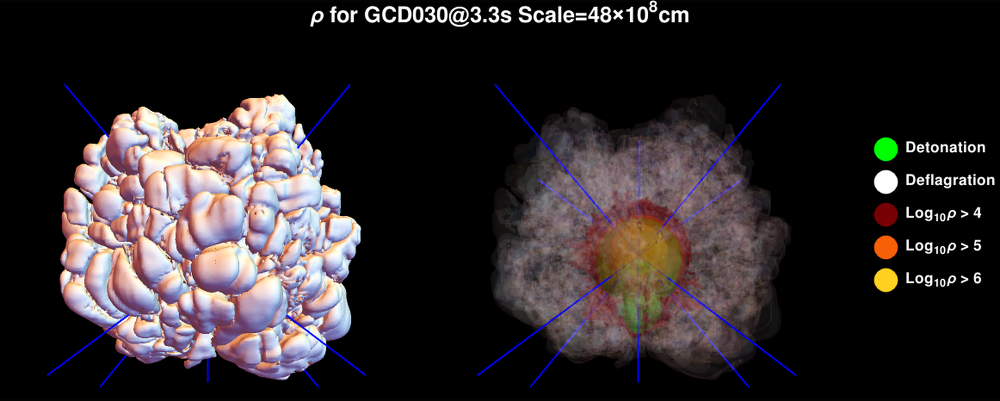} &
\cellcolor{black}\includegraphics[height=0.13\textheight]{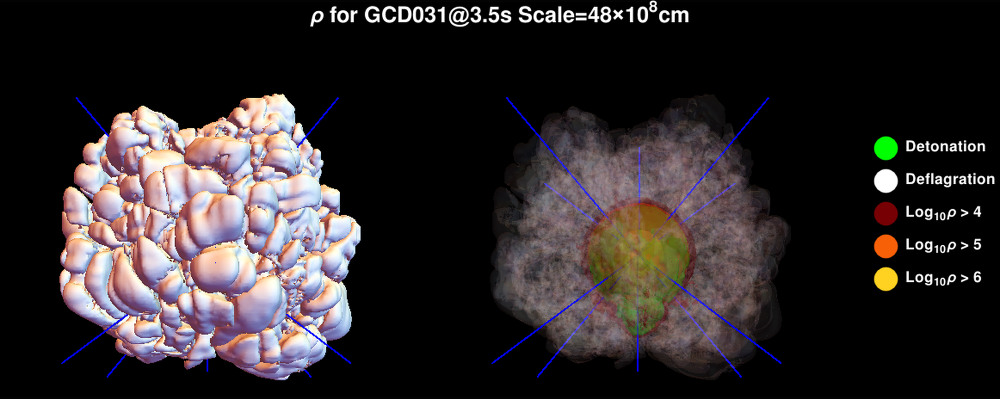} 
\end{tabular}
\caption{Density contour plots including deflagration (white) and detonation (green) surfaces.}
\label{fig:GCD3D}
\end{figure*}
\renewcommand{\arraystretch}{1.5}
\setlength{\tabcolsep}{6pt}

The most significant difference between the two simulations is that for the GCD case the detonation is handled differently because the DDT module is turned off: the detonation is initiated here after 2.37\,s, when the deflagration ashes converge at one point and compress the material to high enough densities and temperatures. The detonation is then initiated by placing an ignition kernel at this spot using a second levelset function to track the detonation front. The similarities between the two simulations make comparing the neutrino signals from each largely dependent on the explosion mechanism (DDT or GCD) and not as dependent on the other simulation details. This is useful considering the high computational cost of these calculations. For the reader's reference, the GCD simulation we use is the case where the initial stellar central density is $2.9\times 10^9\; \text{g cm}^{-3}$ with a temperature of $5\times10^5$ K and the composition is taken to be carbon and oxygen with an initial $Y_e=0.49886$.

\onecolumngrid

\begin{figure}[t]
\includegraphics[trim={0 0 0 1.35cm},clip,width=0.75\linewidth]{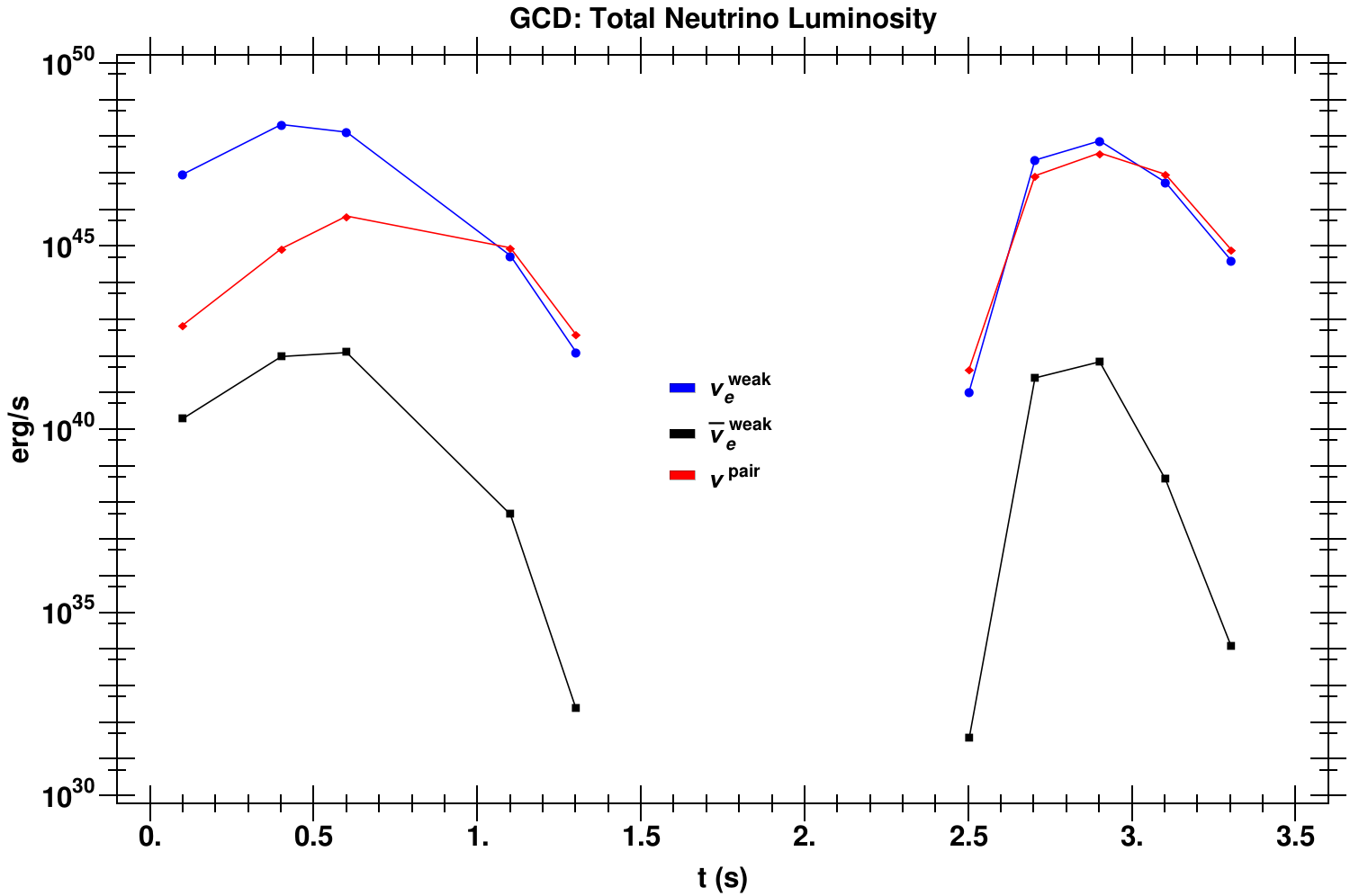}
\caption{GCD total neutrino luminosity for electron neutrino and electron anti neutrino weak processes, and for pair production (all flavors). The dots represent the actual calculation and the lines are a linear interpolation. The middle gap in neutrino emission results from a lack of any NSE stellar zones for these times.}
\label{fig:TotalNeutrinoLuminosityVStime}
\end{figure}
\twocolumngrid

The progression of the SN is depicted in Fig.~\ref{fig:GCD3D}. This figure consists of 14 subplots, each of which contain a title detailing the time since deflagration and the scale of the subplot (length of the blue `axes'). Each subplot also has a legend describing the density contours shown in the two stellar density contour-plots to the left of the legend. In each subplot, the left contour-plot has the white deflagration contour set to opaque to view the exterior development and the right contour-plot has the white deflagration contour set to semi-transparent to allow visualization of the interior evolution. From 2.5\,s onward, the detonation contour is plotted in green. In all subplots, the 10 blue radial lines represent the ten neutrino trajectories that we examine in order to determine the line-of-sight dependence.
One of these trajectories aligns with the initial plume of deflagration burning while another aligns with the point of detonation on the surface. Fig.~\ref{fig:GCD3D} clearly shows how the initial deflagration ash rises to the surface, spreads around the star, collides opposite to the initial emergence point, and how a detonation is initiated there, which spreads quickly throughout the star. Initially only neutrinos which propagate along the plume are affected by the unfolding explosion. The effects of the explosion quickly move to five of the ten trajectories, then nine of ten, and finally all the trajectories we consider are affected. Once the detonation begins, the sequence of affected rays repeats but in the opposite order: the ray last affected during the deflagration phase is the first to be affected by the detonation. The effects of the detonation then spread to the five rays last affected by detonation, then nine rays out of the ten, and finally all ten. Thus we learn the GCD is a highly asymmetrical explosion and while the neutrinos may be emitted more-or-less isotropically because the matter is not sufficiently dense to trap them, we might expect significant viewing-angle dependence on the neutrino signal due to the density profile dependence of neutrino flavor transformation. The detectability of this dependence is explored in Sec. \ref{sec:NeutrinoDetection}. 

Finally, we should point out there is an important difference between the material burnt under deflagration conditions compared to that burnt during the detonation. The deflagration naturally produces turbulence whereas the detonation does not. One of the effects of turbulence is to create a complicated structure for the location of the deflagration flame fronts hence a highly structured profile of density discontinuities. Furthermore the location of the turbulent material in the GCD scenario is quite different from that in the DDT scenario. In the GCD, the detonation region is entirely interior to the mostly exterior deflagration region; in the DDT the deflagration region was closer to the center of the star and much of the detonation region was exterior. We shall see the compound density discontinuities in the deflagration ash on the exterior of the GCD supernova will leave an imprint in the neutrinos, but first we consider the luminosity and neutrino spectra emitted from the explosion.

\section{Neutrino Production \label{sec:Production}}

The neutrino production of the GCD simulation is calculated using the same `NSE only' strategy as in Paper I and used by Odrzywolek and Plewa \cite{Odrzywolek2011a}. This strategy means we only include the zones in Nuclear Statistical Equilibrium (NSE) where $T>3\times10^9$ K. As shown in Paper I, these zones account for the bulk of the emission of neutrinos. The emissivity of these zones is calculated using the software package \textsc{NuLib} \cite{Fuller1982,Langanke2000,Langanke2003,Burrows2006a,Oda1994,Steiner2013,OConnor2010,Sullivan2015}. The weak processes included in our calculation are electron and positron capture on free protons, neutrons, and nuclei. The only thermal process included is electron-positron pair annihilation, which is the dominant emission process for the stellar conditions under examination.

\onecolumngrid

\begin{figure}[t]
\includegraphics[trim={0 2cm 0 1.2cm},clip,width=0.75\linewidth]{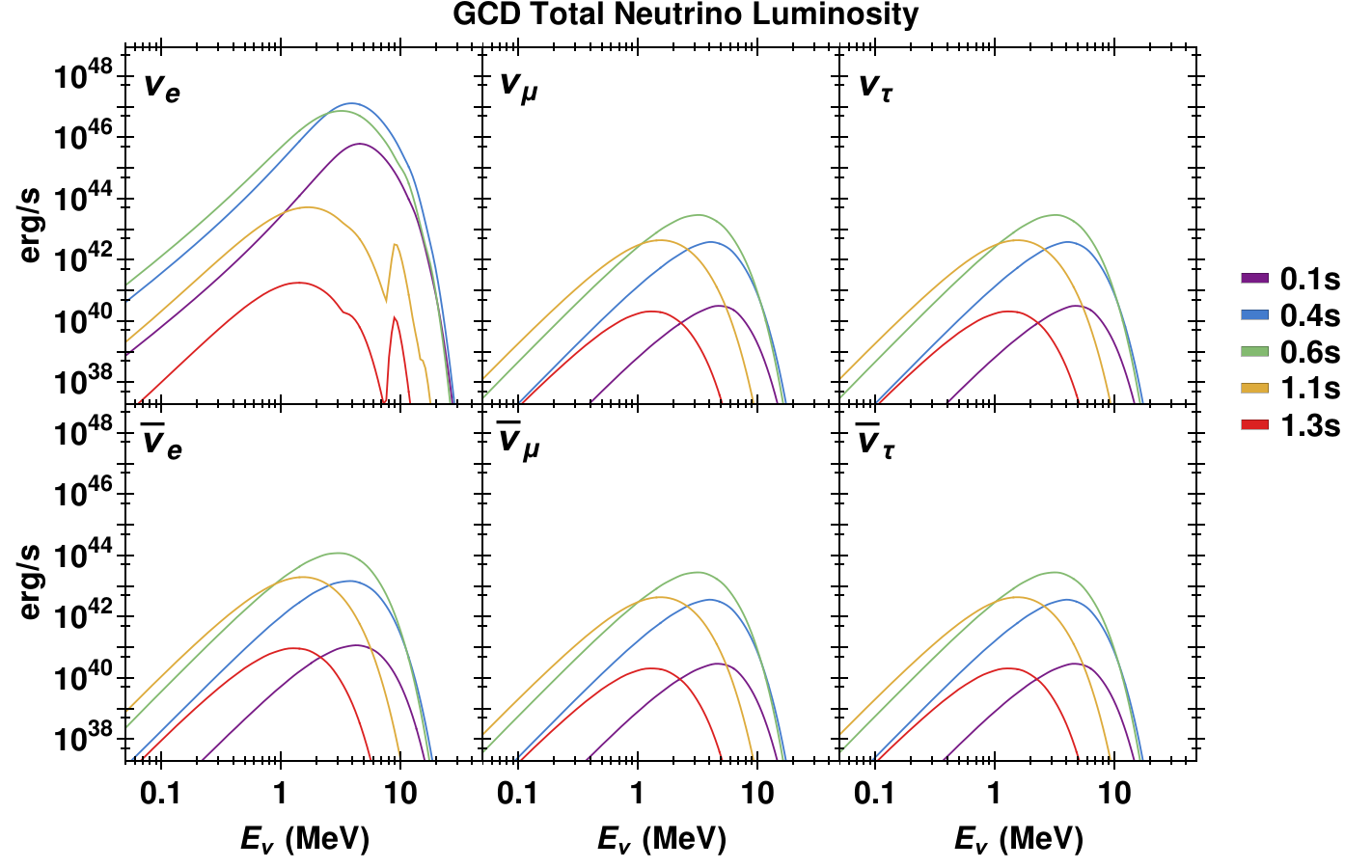}
\includegraphics[trim={0 0 0 1.2cm},clip,width=0.75\linewidth]{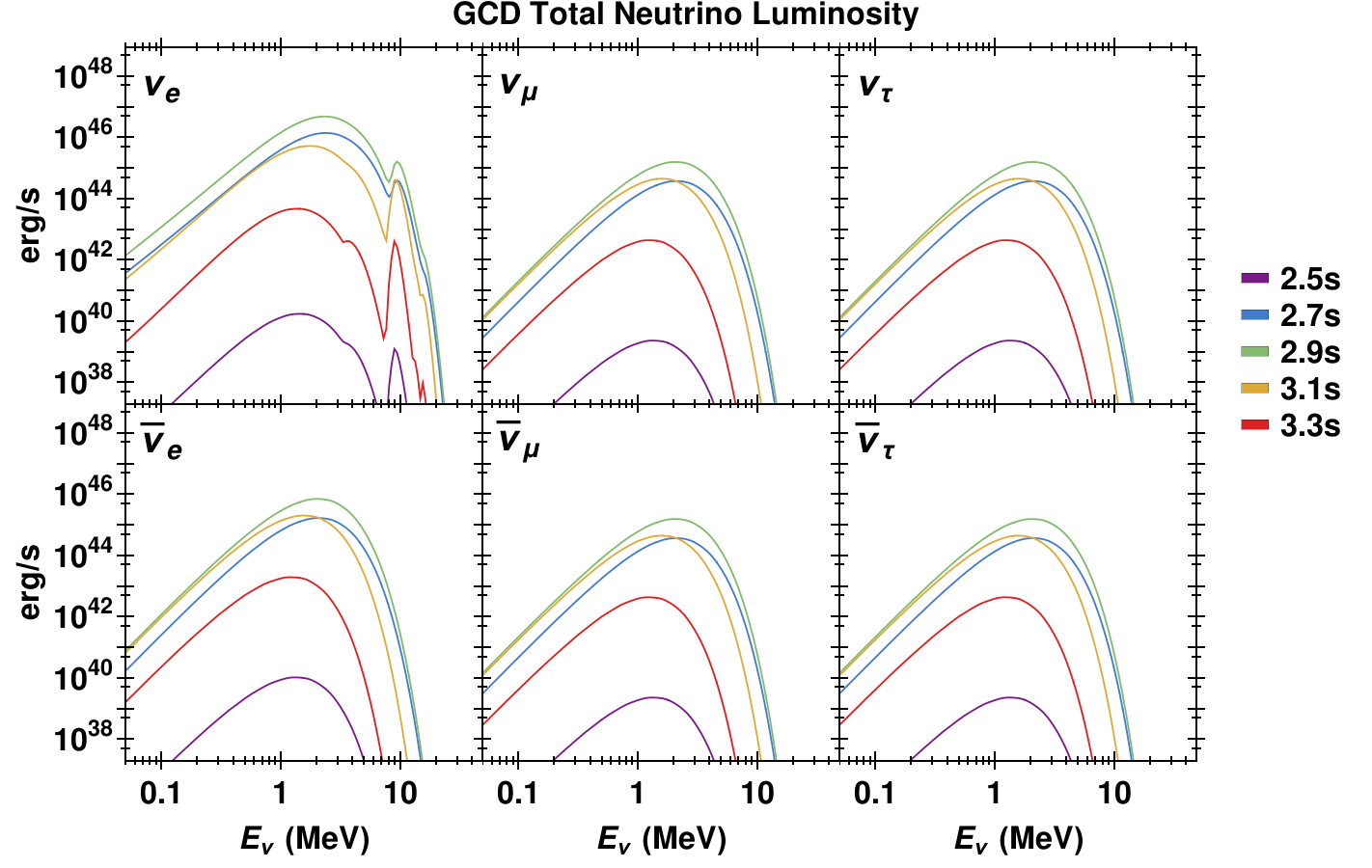}
\caption{GCD neutrino luminosity spectra. Each curve is the sum of all considered weak and thermal processes. The top plots represent the initial ($0.1\text{ s}<t<1.3\text{ s}$) neutrino burst corresponding to the initial deflagration and the bottom plots represent the secondary ($2.5\text{ s}<t<3.3\text{ s}$) neutrino burst corresponding to the detonation.}
\label{fig:NeutrinoLuminositySpectra}
\end{figure}

\begin{figure}[t]
\includegraphics[trim={0 0 0 1.2cm},clip,width=0.75\linewidth]{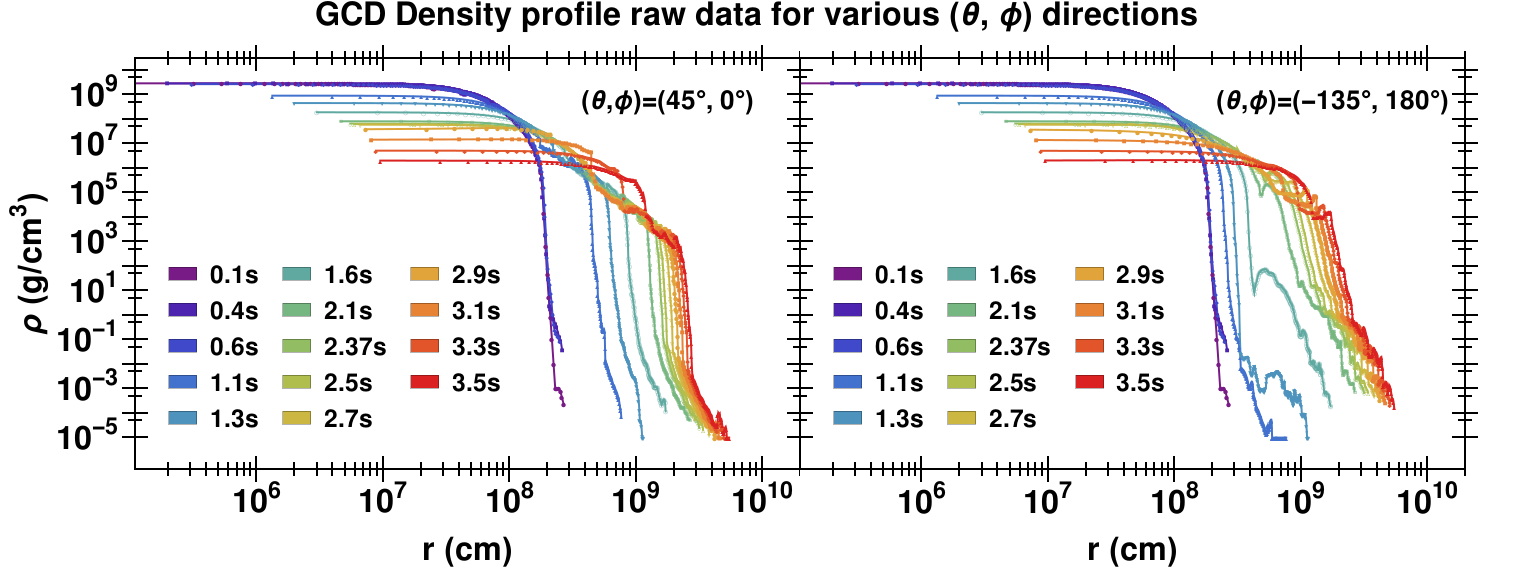}
\caption{Density profiles for various time slices of the GCD simulation. The left panel represents the trajectory that is aligned with the initial deflagration plume (positive z axis) . The right panel represents the trajectory opposite to the left panel's trajectory, i.e. the trajectory aligned with where detonation is initiated (negative z axis).}
\label{fig:Density}
\end{figure}
\twocolumngrid

Figure \ref{fig:TotalNeutrinoLuminosityVStime} shows the neutrino luminosity as a function of time for the GCD simulation. The blue (black) lines and dots represent the $\nu_e$($\bar{\nu}_e$) luminosity arising due to weak processes and the red lines and dots represent the luminosity in all six flavors arising due to pair production. 
We observe from the figure that the neutrino emission from a GCD scenario occurs in two bursts both lasting about $1 \;{\rm s}$ with a gap of around $1.5\;{\rm s}$ between them. The reason for the lack of neutrino emission between $\sim1.3$ s and $\sim2.5$ s - which corresponds to the period when the deflagration spreads over the surface of the star - is because we find the simulation does not contain any zones hot enough to be in NSE during this period. The double burst of neutrinos is a characteristic feature of the GCD explosion mechanism. The neutrinos are produced under very different conditions during the two bursts. During the first burst the material is burning under deflagration conditions and so we shall refer to it as the deflagration neutrino burst; during the second the neutrinos are produced by matter burning under detonation conditions so we shall refer to this second burst as the detonation burst. Despite the different burning conditions, the figure indicates the overall neutrino luminosity during the two bursts are very similar. Closer inspection does reveal a difference: the figure indicates the contribution from thermal processes only begins to be important towards the end of the deflagration burst whereas it is important for the entirety of the detonation burst. 
Finally, the figure indicates the $\bar{\nu}_e$ luminosity arising due to weak processes is vastly superseded by the $\bar{\nu}_e$ luminosity arising due to thermal processes. Electron antineutrino emission from weak processes occurs as a consequence of positron capture by neutrons in nuclei and, at these temperatures and densities, these reactions are energetically disfavored. 

\IRS{Our results are similar to the findings of Odrzywolek \& Plewa for their Y12 GCD model, with a few differences. The total energy radiated by neutrinos in our model is about a factor of six lower compared to their result, mostly due to the fact that their model burns more mass during the deflagration. Since most of the neutrino luminosity is due to electron captures on free protons during the deflagration phase, the amount of mass burned in the deflagration is the most important factor in determining the total energy carried by neutrinos. The main reason why the Y12 model burns more mass in the deflagration is that the simulations are performed in 2D with assumed azimuthal rotational symmetry, unlike the GCD200 model, which is a full 3D simulation. With this 
imposed symmetry, deflagration bubbles in 2D away from the z-axis effectively turn into tori, which 
enlarges the volume burned compared to a 3D simulation.}

\IRS{We also observe the detonation burst occurs about ${\sim}1\;\text{s}$ earlier in our simulation than in the Y12 model of Odrzywolek \& Plewa. We attribute this difference largely to the different assumed ignition conditions, in particular the initial offset of the deflagration spark. The deflagration of the Y12 model evolves and rises much more slowly compared to the GCD200 model. The Rayleigh-Taylor (RT) growth rate scales as the square root of the local acceleration due to gravity. Our GCD200 deflagration is ignited at an offset distance of 200 km from the center, whereas the Y12 ignition occurs a mere 12.5 km offset from the center. The closer the ignition to the center, the smaller the gravity and the slower the growth-rate of the RT-instability and hence the slower the evolution of the deflagration. The fact that their model uses a lower central density $2.0\times10^9\,\mathrm{g}\,\mathrm{cm}^{-3}$ vs $2.9\times10^9\,\mathrm{g}\,\mathrm{cm}^{-3}$) further amplifies the difference.  We have chosen a 200\,km offset because our 3D simulations have shown that smaller offsets do not lead to detonations \citep{Seitenzahl2016}.}


The spectral features associated with the GCD SN are presented in Fig. \ref{fig:NeutrinoLuminositySpectra}. The top six plots represent the luminosity for the six neutrino flavors for the five time slices corresponding to the initial deflagration burning phase. The bottom six plots represent the luminosity for the six neutrino flavors for the five time slices corresponding to the final detonation burning. Figure \ref{fig:NeutrinoLuminositySpectra} indicates that  initially the $\nu_e$ luminosity is much larger than the luminosity in other flavors but by midway through the deflagration burst the luminosity in all flavors is similar and remains this way for the later detonation burst. We also observe in the figure a spectral feature at $\sim10$ MeV (mostly from electron capture on copper \cite{Wright2016}) which is absent initially in the deflagration burst but then emerges at $t\sim1$ s and remains present in the spectrum for the entirety of the detonation burst. 

\section{Neutrino Flavor Transformation\label{sec:NeutrinoOscillation}}

\begin{figure*}[t]
\includegraphics[trim={0 0 0 4cm},clip,width=0.8\linewidth]{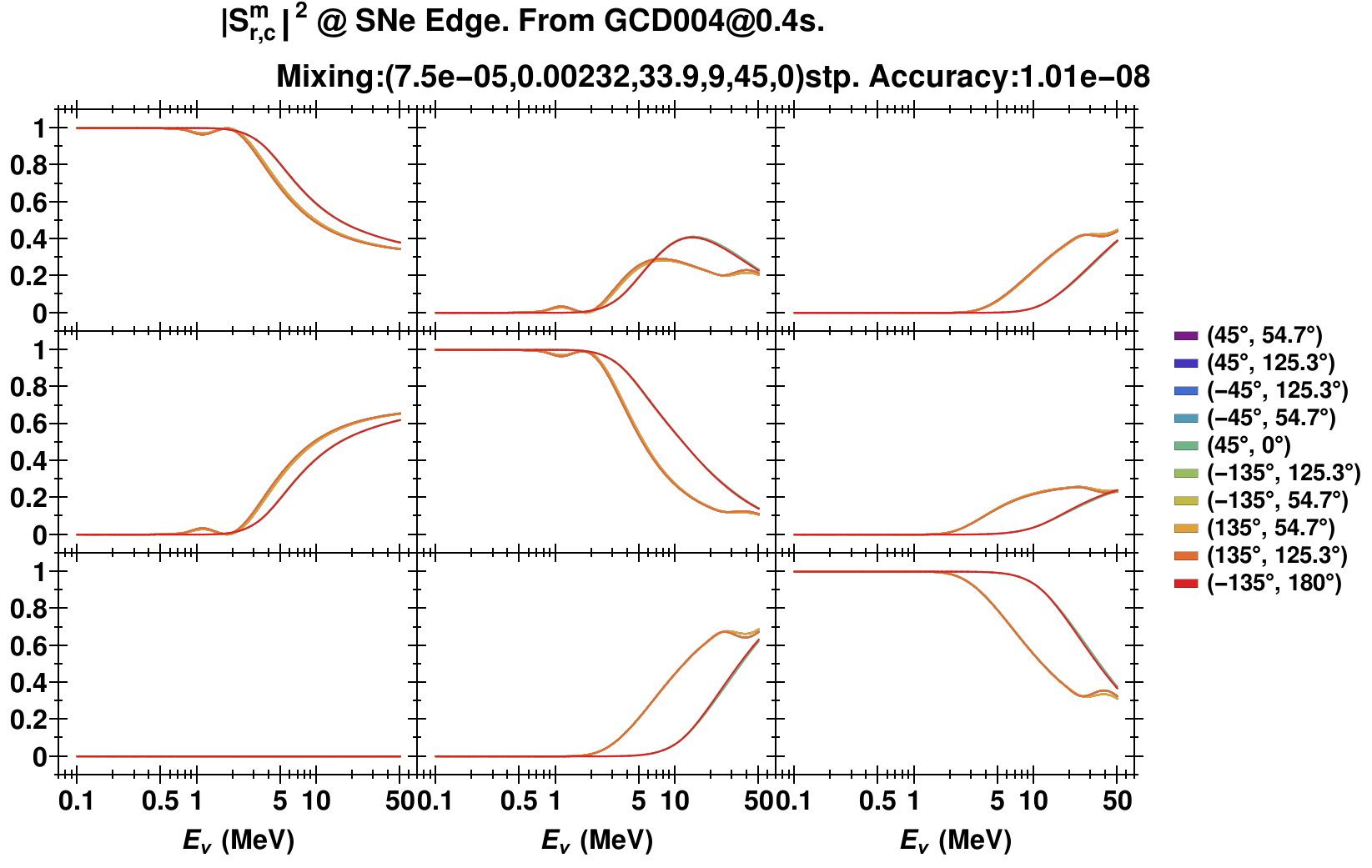}
\caption{The neutrino matter basis transition probabilities $P^\text{(m)}_{ij}\left(E_\nu\right)$ as a function of energy. The top row, from left to right, shows $P^\text{(m)}_{11}$, $P^\text{(m)}_{12}$ and $P^\text{(m)}_{13}$. The middle row from left to right shows $P^\text{(m)}_{21}$, $P^\text{(m)}_{22}$ and $P^\text{(m)}_{23}$ and the bottom row from left to right shows $P^\text{(m)}_{31}$, $P^\text{(m)}_{32}$ and $P^\text{(m)}_{33}$. The mass ordering is normal and the snapshot time is $t=0.4\;\text{s}$. The polar and azimuth angles of each trajectory are given in the legend.}
\label{fig:SmNHAt0.4s}
\end{figure*}

\begin{figure*}[t]
\includegraphics[trim={0 0 0 4.2cm},clip,width=0.8\linewidth]{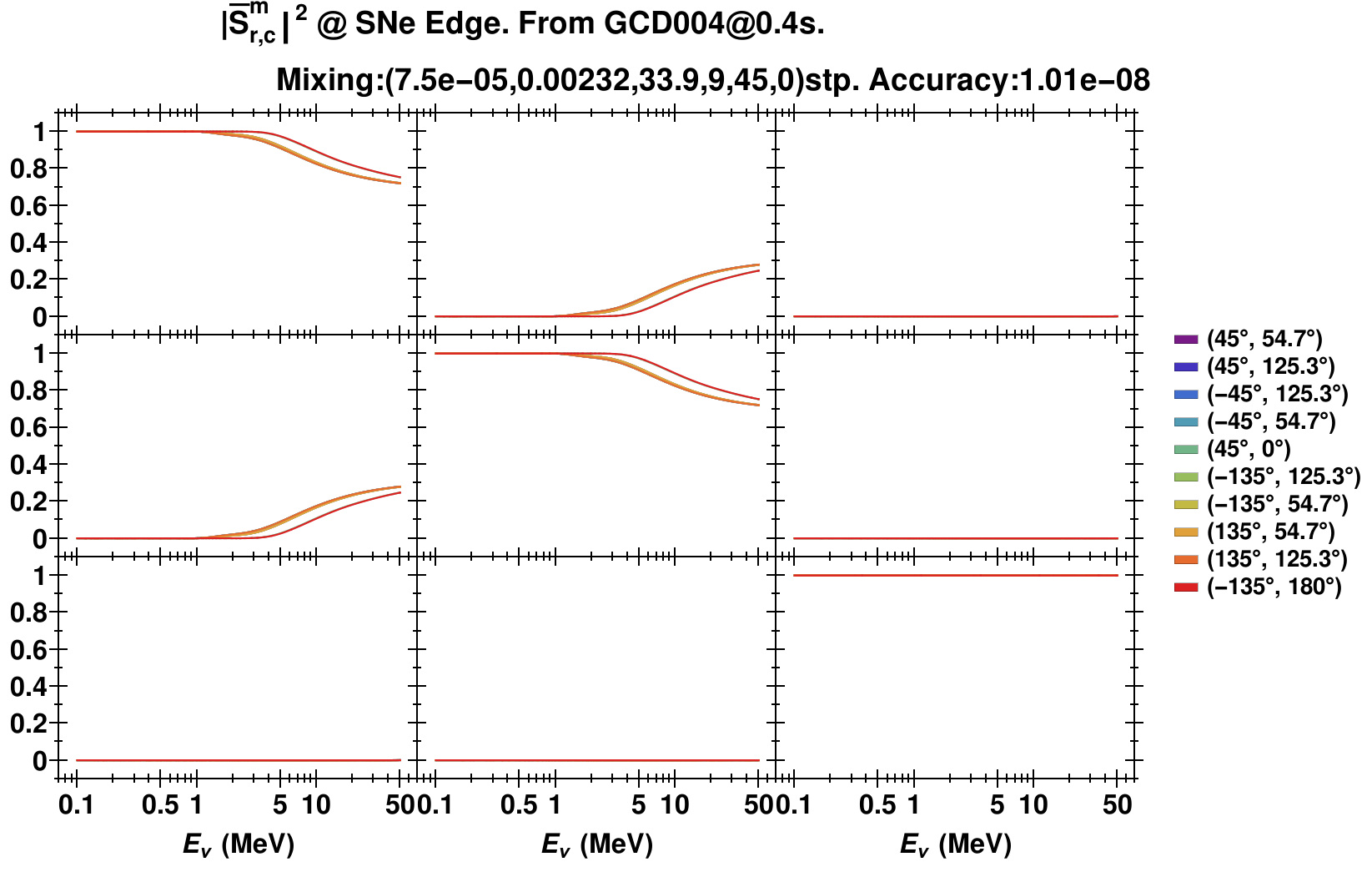}
\caption{The antineutrino matter basis transition probabilities for normal mass ordering at \textbf{$t=0.4\;\text{s}$}. The structure and layout is the same as that of Fig. \ref{fig:SmNHAt0.4s}.}
\label{fig:SmBarNHAt0.4s}
\end{figure*}

\begin{figure*}[t]
\includegraphics[trim={0 0 0 4cm},clip,width=0.8\linewidth]{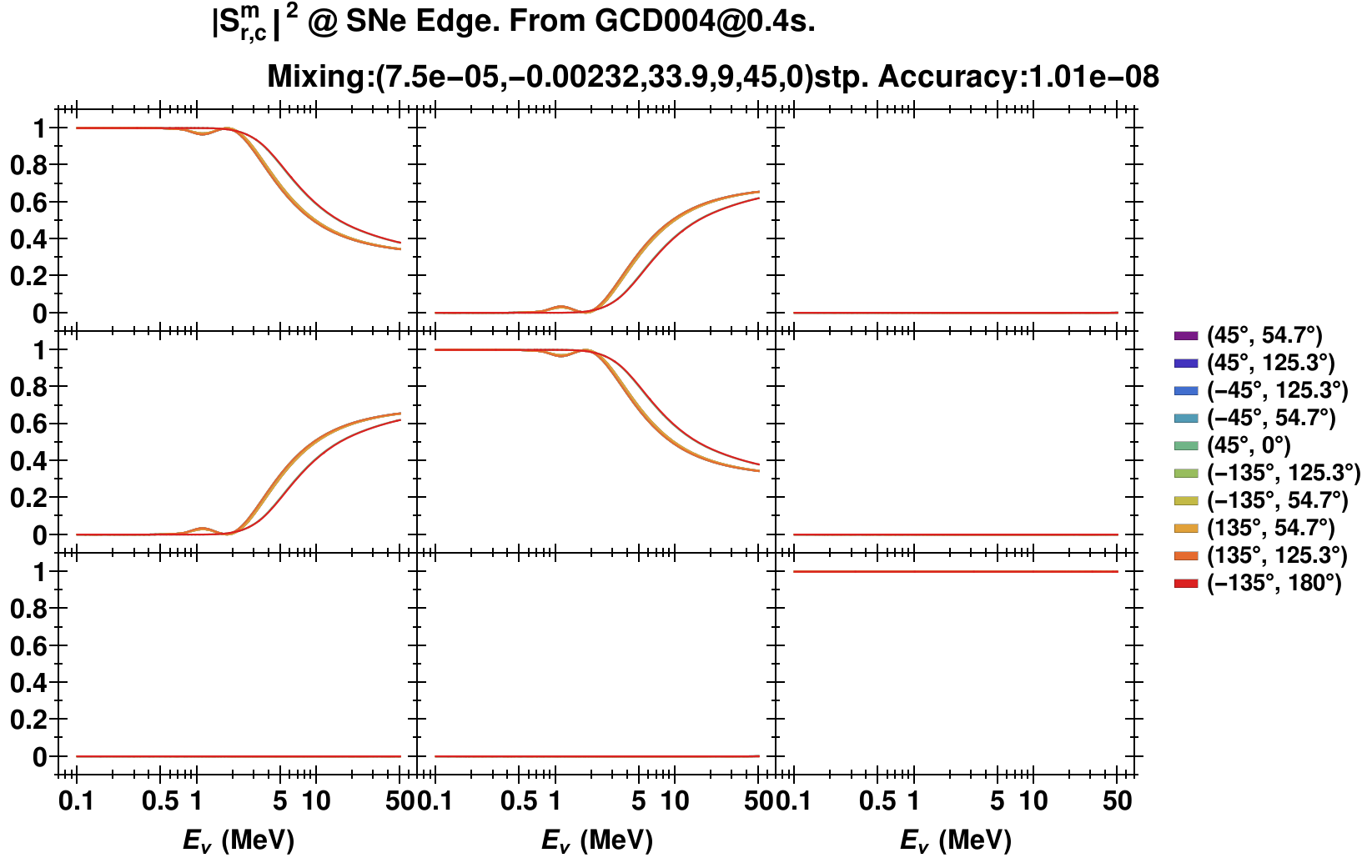}
\caption{The neutrino matter basis transition probabilities for inverted mass ordering at \textbf{$t=0.4\;\text{s}$}. The structure and layout is the same as that of Fig. \ref{fig:SmNHAt0.4s}.}
\label{fig:SmIHAt0.4s}
\end{figure*}

\begin{figure*}[t]
\includegraphics[trim={0 0 0 4.2cm},clip,width=0.8\linewidth]{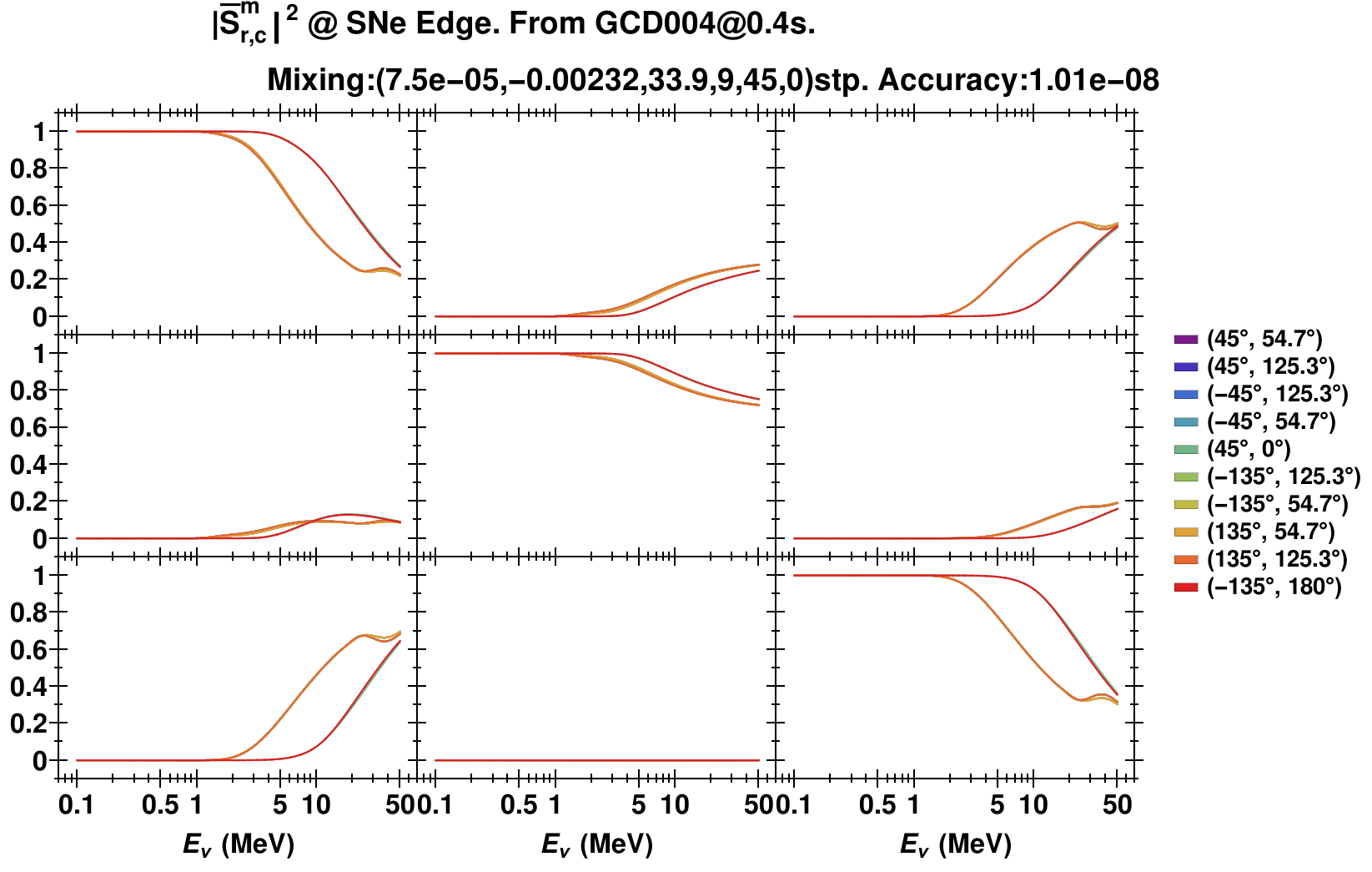}
\caption{The antineutrino matter basis transition probabilities for inverted mass ordering at \textbf{$t=0.4\;\text{s}$}. The structure and layout is the same as that of Fig. \ref{fig:SmNHAt0.4s}.}
\label{fig:SmBarIHAt0.4s}
\end{figure*}

The neutrino flavor composition of the spectra emitted in the supernova are not what will be detected at Earth due to the flavor transformations as the neutrinos propagate through the supernova and due to decoherence on the journey to Earth. This has the potential to mix the time and spectral features of one flavor with those of another. The flavor transformations are dependent on the density and electron fraction of the material through which the neutrinos must propagate \cite{1978PhRvD..17.2369W,Mikheyev:1985aa}. The density profiles along two lines of sight through the simulation are shown in Fig. \ref{fig:Density}: the density profiles along the deflagration plume (left panel) and its polar opposite along the ray that passes through the detonation point (right panel). This density plot can reveal the effects of turbulence as departures from the smooth density profile seen at $t=0.1\text{ s}$. The turbulent effects in the high density deflagration ash during the early deflagration phase are shown in the leftmost panel for $2\times10^8<r\text{ (cm)}<5\times10^8$. While the turbulent effects in the low density deflagration ash during the later detonation phase are shown in the rightmost panel for $4\times10^8<r\text{ (cm)}<2\times10^9$.
Along both rays, the nontrivial density profiles mean we cannot assume adiabatic neutrino flavor propagation. Instead, the flavor transition probabilities are computed numerically following the same protocol as Paper I. Since the calculation and many of the features in the results are the same as in Paper I, in what follows we just present the results of the transformation calculations.

\subsection{Matter basis transformation probabilities\label{sec:PMatter}}
The neutrino transformation calculations are undertaken and presented in the matter basis in order to remove the trivial mixing in the flavor basis. 
As a reference, in dense matter and the normal mass ordering (NMO) the electron neutrino flavor is closely aligned with the matter basis state $\nu_3$ and the electron antineutrino with matter basis state $\bar{\nu}_1$. For an inverted mass ordering (IMO) the electron flavor in dense matter is closely aligned with the matter basis state $\nu_2$ and the electron antineutrino with matter basis state $\bar{\nu}_3$. In vacuum, the electron flavor neutrino is approximately $70\%$ matter state $\nu_1$ and $30\%$ matter state $\nu_2$, the electron antineutrino is $\sim 70\%$ matter state $\bar{\nu}_1$ and $30\%$ matter state $\bar{\nu}_2$.

\begin{figure*}[t]
\includegraphics[trim={0 0 0 4cm},clip,width=0.8\linewidth]{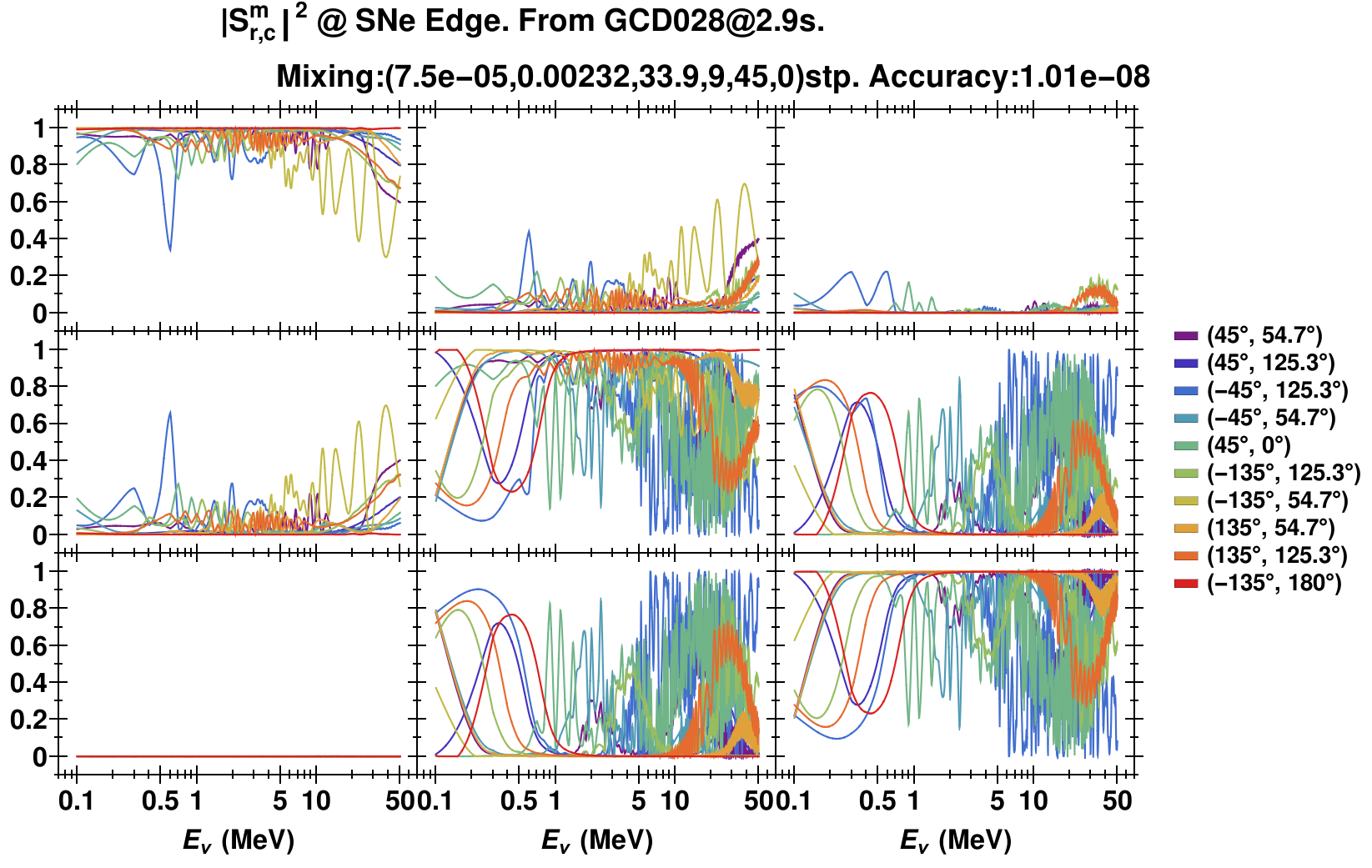}
\caption{The neutrino matter basis transition probabilities for normal mass ordering at \textbf{$t=2.9\;\text{s}$}. The structure and layout is the same as that of Fig. \ref{fig:SmNHAt0.4s}.}
\label{fig:SmNHAt2.9s}
\end{figure*}

\begin{figure*}[t]
\includegraphics[trim={0 0 0 4.2cm},clip,width=0.8\linewidth]{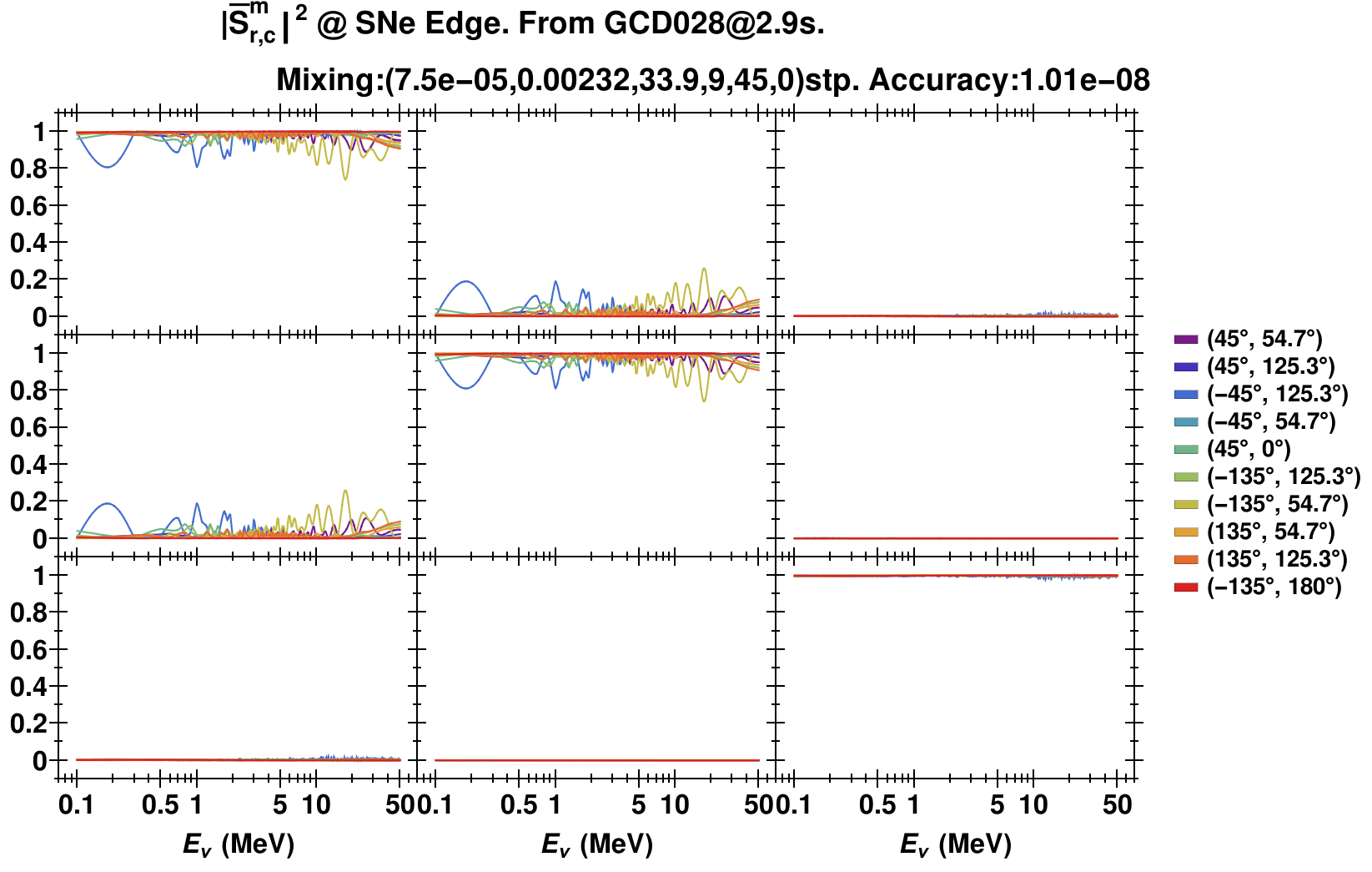}
\caption{The antineutrino matter basis transition probabilities for normal mass ordering at \textbf{$t=2.9\;\text{s}$}. The structure and layout is the same as that of Fig. \ref{fig:SmNHAt0.4s}.}
\label{fig:SmBarNHAt2.9s}
\end{figure*}

The matter basis transition probabilities $P^{\text{(m)}}_{ij} = P(\nu^{\text{(m)}}_j \rightarrow \nu^{\text{(m)}}_i)$ for a neutrino that was initially state $j$ in the matter basis to be detected as state $i$ in the matter basis after the neutrinos have propagated through the SN depend upon the the neutrino mass orderings, the line of sight and the epoch. We have computed the transition probabilities at the ten different snapshot times indicated by the points in Fig. \ref{fig:NeutrinoLuminositySpectra} but here show just the results at the two points of maximum luminosity, namely, at the peak of the deflagration neutrino burst at $t = 0.4$ s and at the peak of the detonation neutrino burst at $t = 2.9$ s. 

\begin{figure*}[t]
\includegraphics[trim={0 0 0 4cm},clip,width=0.8\linewidth]{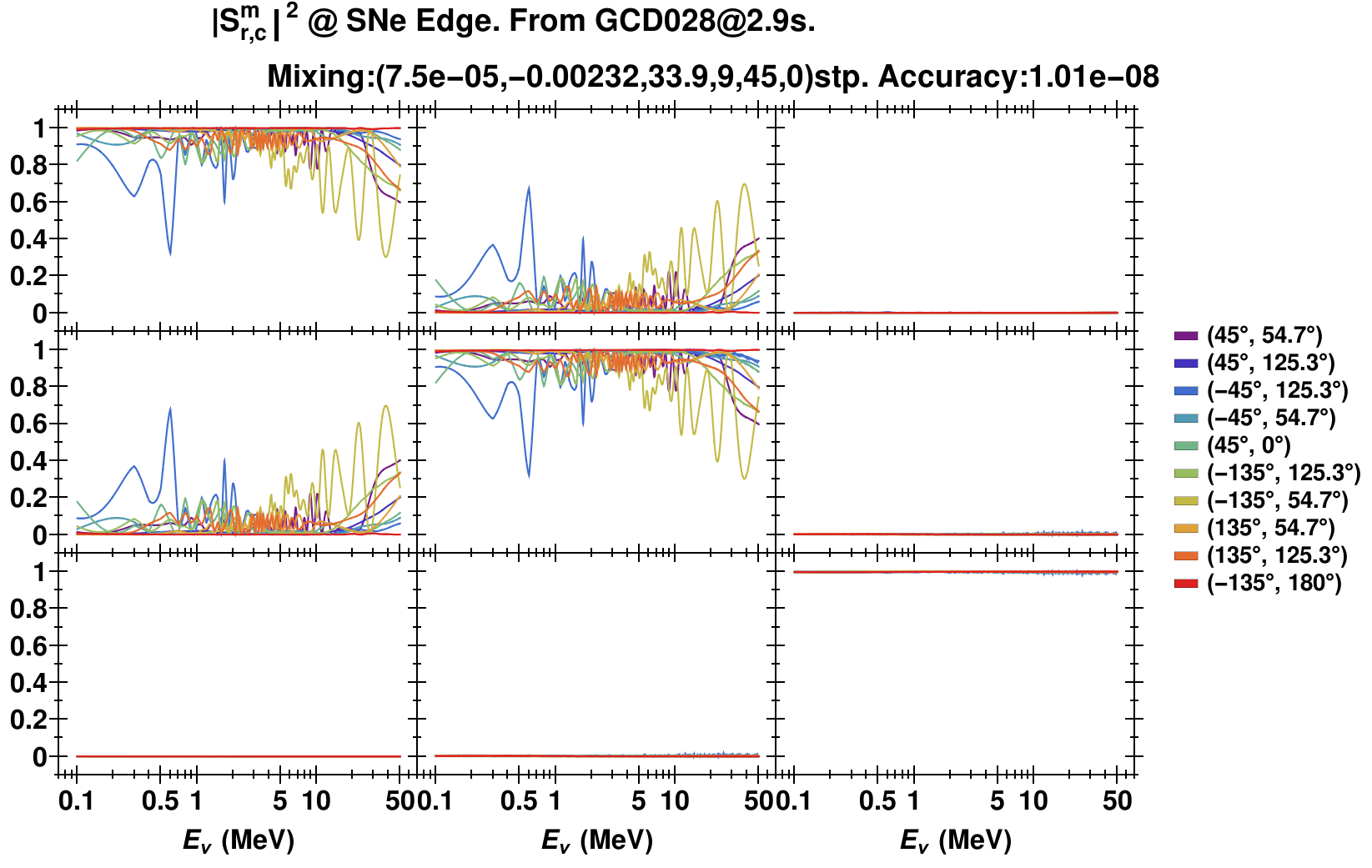}
\caption{The neutrino matter basis transition probabilities for inverted mass ordering at \textbf{$t=2.9\;\text{s}$}. The structure and layout is the same as that of Fig. \ref{fig:SmNHAt0.4s}.}
\label{fig:SmIHAt2.9s}
\end{figure*}

\begin{figure*}[t]
\includegraphics[trim={0 0 0 4.2cm},clip,width=0.8\linewidth]{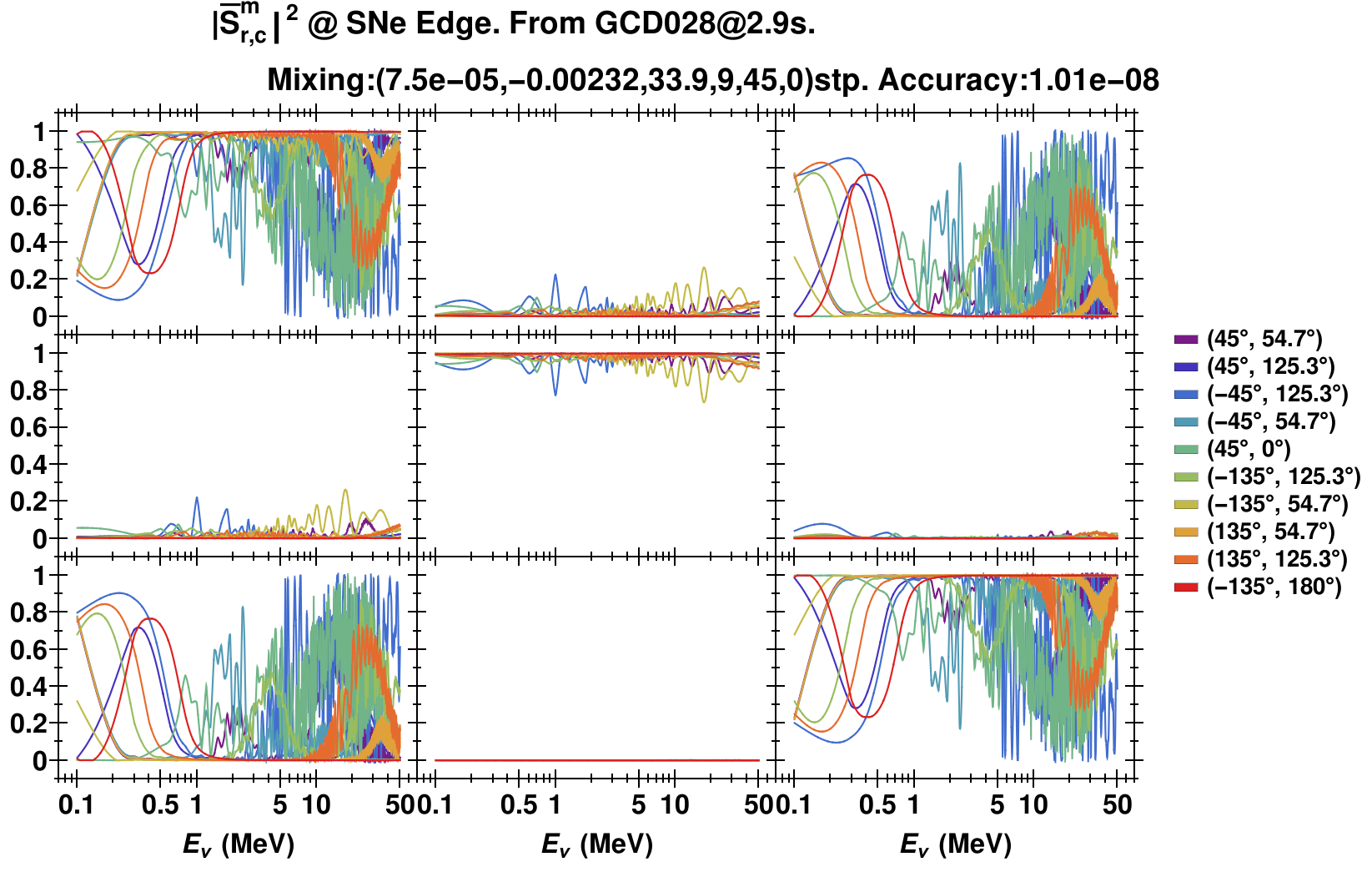}
\caption{The antineutrino matter basis transition probabilities for inverted mass ordering at \textbf{$t=2.9\;\text{s}$}. The structure and layout is the same as that of Fig. \ref{fig:SmNHAt0.4s}.}
\label{fig:SmBarIHAt2.9s}
\end{figure*}

At the deflagration peak ($t=0.4$ s), Fig. \ref{fig:SmNHAt0.4s} shows that for $E_\nu\lesssim1$ MeV, the neutrino transition probabilities in the NMO are adiabatic ($P_{ij} \sim \delta_{ij}$) for all ten lines of sight considered. However, for $E_\nu>1$ MeV, the neutrino transition probabilities in the NMO become partially diabatic. The diabaticity is a consequence of the very steep density profile at the edge of the star at this epoch. The transition probabilities also exhibit line-of-sight dependence since the transition probability versus energy curves in the figure depend upon the exact trajectory the neutrinos take through the star. The line-of-sight dependence at this epoch is almost binary and the curves fall into two groups depending upon whether or not the neutrinos have to traverse the deflagration plume. The effect is substantial with a line-of-sight variation as big as $50\%$ difference for $E\sim10$ MeV in the  $P^{\text{(m)}}_{33}$ channel. This change in $P^{\text{(m)}}_{33}$ can have consequences for the signal, because for a NMO, the electron flavor neutrinos are closely aligned with matter state $\nu_3$ in dense matter whereas matter state $\nu_3$ in the vacuum has very little electron flavor content.

\begin{figure*}[t]
\includegraphics[trim={0 0 0 2.2cm},clip,width=0.75\linewidth]{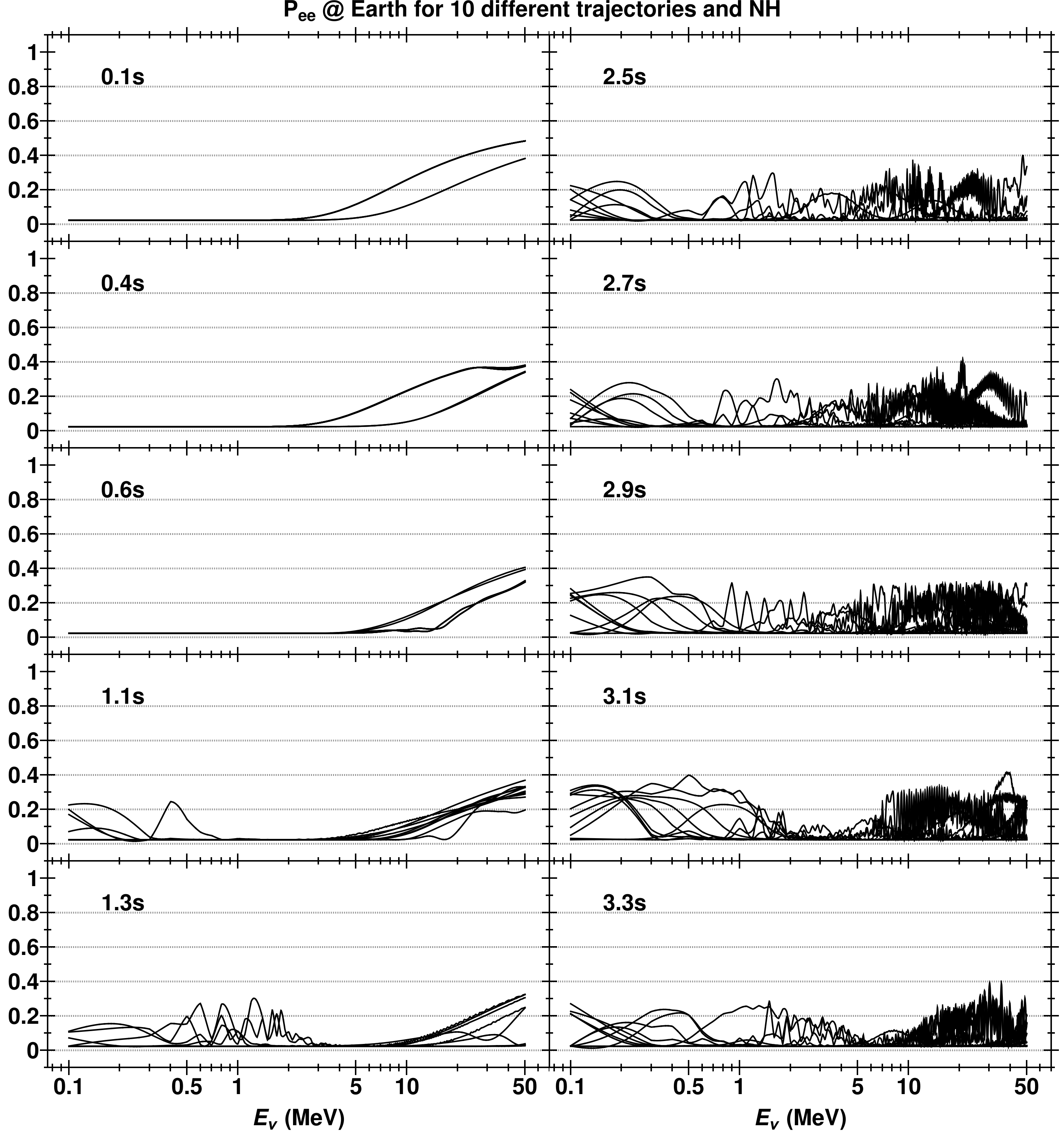}
\caption{The electron neutrino survival probability $P_{ee}$ at Earth as a function of neutrino energy for each of the ten lines-of-sight through the simulation at the ten snapshot times. The mass ordering is normal.}
\label{fig:PeeEarthTblNH}
\end{figure*}

At the same epoch Fig. \ref{fig:SmBarNHAt0.4s} shows the transition probabilities for the antineutrinos again in the NMO. 
For this mass ordering the propagation is adiabatic except in the $\bar{\nu}_1 \leftrightarrow \bar{\nu}_2$ channel 
where we find some mixing occurs for energies above $E_\nu\gtrsim1$ MeV which depends upon the line of sight. 
For this mass ordering and at this epoch, the amount of mixing and the line-of-sight variability in the antineutrinos is not as large as the mixing in the same channel for the neutrinos. 

That situation changes if we swap the neutrino mass ordering. For the IMO the neutrino and antineutrino transition probabilities at the peak of the deflagration burst are shown in Figs. \ref{fig:SmIHAt0.4s} and Fig. \ref{fig:SmBarIHAt0.4s}. 
Now the mixing channels involving $\nu_3$ are totally adiabatic with zero line-of-sight dependence. The mixing between 
$\nu_1$ and $\nu_2$ is very similar to the NMO with adiabatic propagation up to $E_{\nu} \sim 5\;{\rm MeV}$ and partially adiabatic thereafter. 
The line-of-sight dependence in the IMO is only a difference in the transition probabilities of $\sim 10\%$ at most. 
For the antineutrinos and the IMO we also observe adiabaticity in all channels up to $E_{\nu} \sim 5\;{\rm MeV}$ with partial adiabatic evolution for high energies. As seen previously in other mixing scenarios, at this epoch the transition probabilities as a function of energy can be grouped into one of two groups depending upon whether the deflagration plume was crossed. 
As for the NMO, the fact that the matter basis probabilities are not adiabatic has consequences for the signal we expect. For an IMO the electron flavor in dense matter aligns with matter state $\nu_2$: in vacuum matter state $\nu_2$ is only $30\%$ electron flavor. Conversion of matter state $\nu_2$ to $\nu_1$ which is $70\%$ electron flavor in vacuum boosts the observability of the original electron neutrino spectrum by more than double. 


\begin{figure*}[t]
\includegraphics[trim={0 0 0 2.7cm},clip,width=0.75\linewidth]{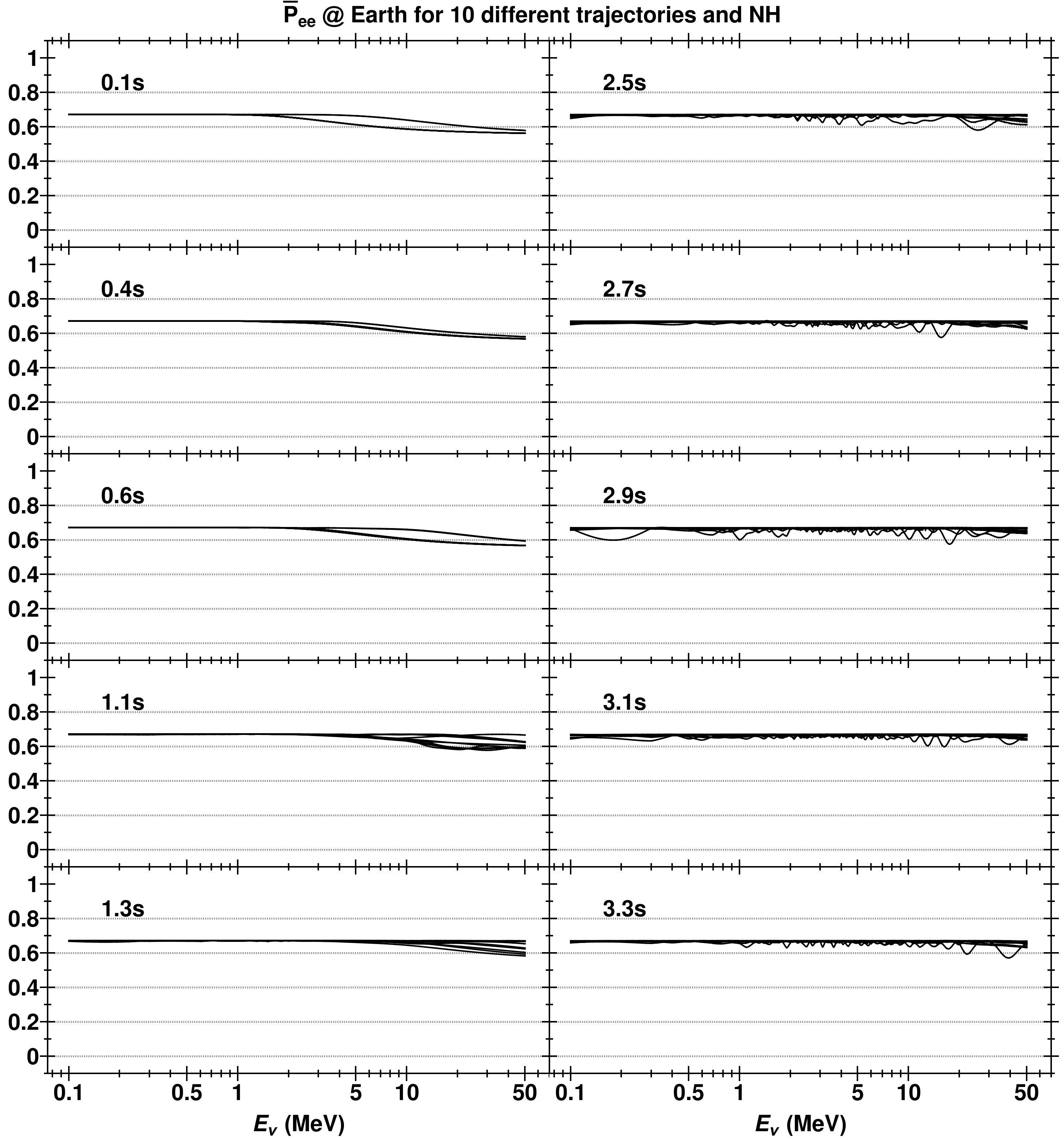}
\caption{The electron antineutrino survival probability $\bar{P}_{ee}$ at Earth as a function of neutrino energy for each of the ten lines-of-sight through the simulation at the ten snapshot times. The mass ordering is normal.}
\label{fig:PBareeEarthTblNH}
\end{figure*}

\begin{figure*}[t]
\includegraphics[trim={0 0 0 2.2cm},clip,width=0.75\linewidth]{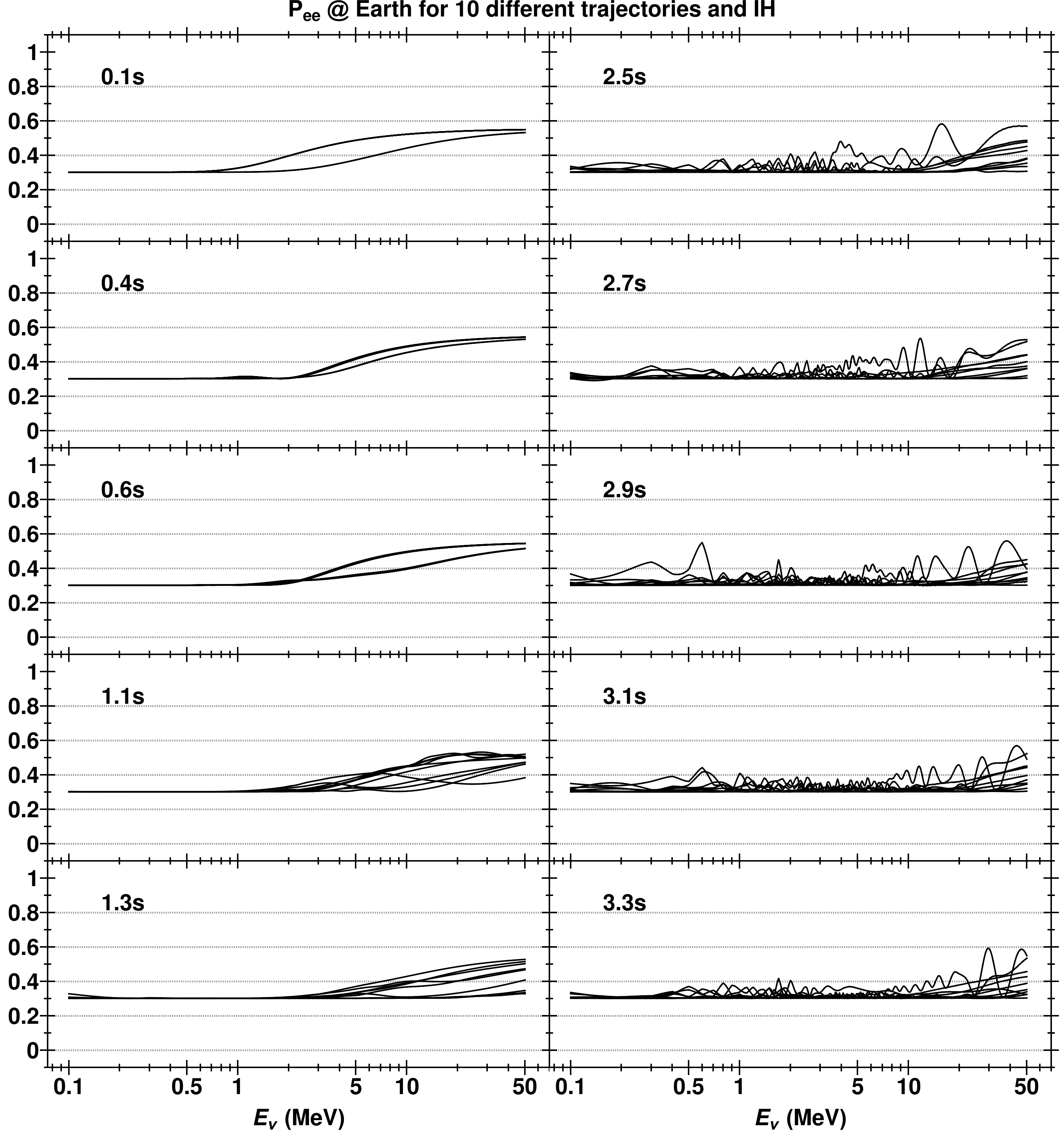}
\caption{The same as figure (\ref{fig:PeeEarthTblNH}) but for an inverted mass hierarchy.}
\label{fig:PeeEarthTblIH}
\end{figure*}

\begin{figure*}[t]
\includegraphics[trim={0 0 0 2.7cm},clip,width=0.75\linewidth]{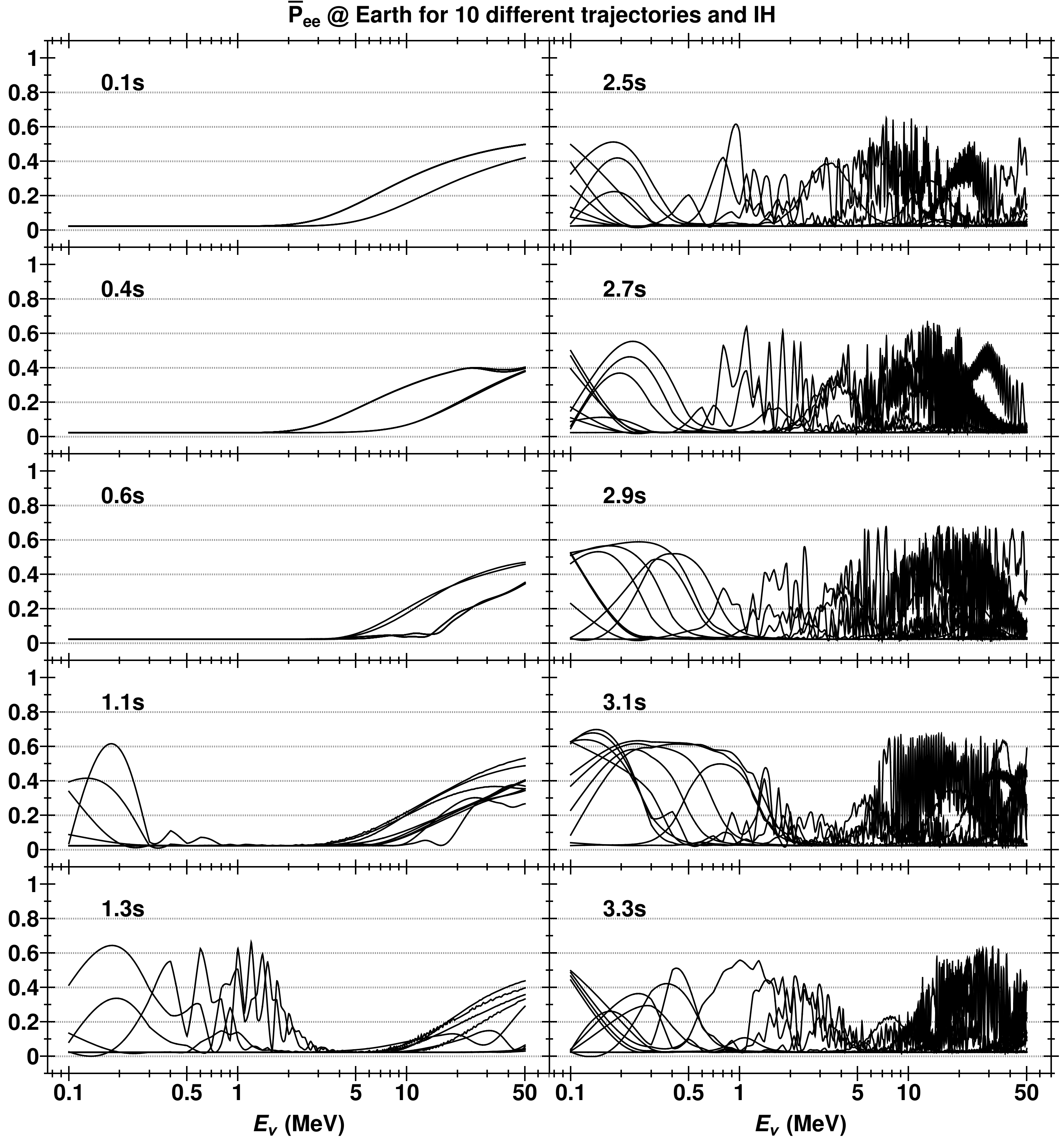}
\caption{The same as figure (\ref{fig:PBareeEarthTblNH}) but for an inverted mass hierarchy.}
\label{fig:PBareeEarthTblIH}
\end{figure*}

The line-of-sight dependence at $t=0.4\;{\rm s}$ is essentially binary. If we consider the later epoch of $t=2.9$ s at the peak of the detonation burst the line-of-sight dependence becomes much more extreme. The transition probabilities in the NMO for the neutrinos and antineutrinos at the detonation burst peak are shown in Figs. \ref{fig:SmNHAt2.9s} and \ref{fig:SmBarNHAt2.9s} respectively. When compared to Figs. \ref{fig:SmNHAt0.4s} and \ref{fig:SmBarNHAt0.4s} we see each line of sight produces a unique curve. The difference from one to the next can be as large as 100\% for the neutrinos and $E\gtrsim 5$ MeV. The lower energy neutrinos and antineutrinos no longer evolve adiabatically in any channel except $P_{31}^{(m)}$. At higher energies, $E\gtrsim 5$ MeV, the variability in $P^\text{(m)}_{22}$, $P^\text{(m)}_{23}$, $P^\text{(m)}_{32}$ and $P^\text{(m)}_{33}$ appears to be much larger than at $t=0.4\;{\rm s}$ but, in contrast, the transition probabilities $P^\text{(m)}_{11}$, $P^\text{(m)}_{12}$, $P^\text{(m)}_{13}$ and $P^\text{(m)}_{21}$ at higher energies are closer to adiabatic than at the previous epoch. The same is also true for the higher energies antineutrinos in this mass ordering at this epoch: the variability between the different lines of sight has increased but the overall trend is towards greater adiabaticity for $E\gtrsim 5$ MeV. This evolution is similar to what was seen in Paper I for the DDT case where it was explained as due to a softening of the density gradients as the star explodes. Figure \ref{fig:Density} indicates there is a similar softening for the GCD scenario. 

Finally, Figs. \ref{fig:SmIHAt2.9s} and \ref{fig:SmBarIHAt2.9s} show the transition probabilities at the detonation peak when we switch to an IMO. Now we observe the mixing involving $\nu_3$ remains adiabatic at all energies while the mixing between $\nu_1$ and $\nu_2$ at $E\lesssim 5$ MeV is less adiabatic but for $E\gtrsim 5$ MeV it is more adiabatic. 
In the antineutrinos, $\bar{P}^{\text{(m)}}_{22}$ is much less affected by the turbulent deflagration ash than $\bar{P}^{\text{(m)}}_{11}$ and $\bar{P}^{\text{(m)}}_{33}$ which are now the channels where strong variability of the transition probabilities is found. 

An analysis of the causes of the large amplitude variability in the matter transition probabilities during the detonation neutrino burst seen in Figs. \ref{fig:SmNHAt2.9s} to \ref{fig:SmBarIHAt2.9s} at $t=2.9\;{\rm s}$ indicates they are due to the aforementioned density discontinuities present in the deflagration ash layer on the `surface' of the star. The neutrinos emitted during the detonation burst are produced in regions which lie interior to the ash and thus must pass through these density discontinuities as they exit the supernova. The densities in the deflagration ash layer located on the surface are similar to the resonance densities of the high (H) and low (L) Mikheyev-Smirnov-Wolfenstein (MSW) resonances \cite{2000PhRvD..62c3007D} for the neutrino energies we consider. This setup is a recipe for the emergence of strong phase effects in the matter transition probabilities - compare with the results from Lund and Kneller \cite{2013PhRvD..88b3008L} for the case of core-collapse neutrinos. The phase effects appears to be greatest in the H resonance channels - $\nu_2 \leftrightarrow \nu_3$ for a NMO, $\bar{\nu}_1 \leftrightarrow \bar{\nu}_3$ for a IMO. A more detailed discussion of the oscillation phenomenology, including turbulent and phase effects, can be found in Paper I \cite{Wright2016}. 

\subsection{Flavor basis transition probabilities at Earth\label{sec:PFlavor}}
Once the matter basis oscillation probabilities are calculated we need to account for decoherence and then perform a basis change in order to compute the flavor basis transition probabilities at Earth. The details of this process are, again, given in Paper I and so here we will again just discuss some of the results. Figures \ref{fig:PeeEarthTblNH} and \ref{fig:PeeEarthTblIH} show $P_{ee}\left(E_\nu\right)$ on Earth for both NMO and IMO respectively. The different black lines in these figures represent the ten neutrino trajectories considered and each subplot is for a different time slice. In a similar fashion, but for antineutrinos, Figs. \ref{fig:PBareeEarthTblNH} and \ref{fig:PBareeEarthTblIH} show $\bar{P}_{ee}\left(E_\nu\right)$ on Earth for both NMO and IMO respectively. These figures reveal that much of the oscillation phenomenology discussed in Sec. \ref{sec:PMatter} for the matter basis at the edge of the supernova is also seen in the flavor basis at Earth. In particular, (1) At all epochs the effect of diabatic propagation is to increase the electron flavor survival probability from $\sim 0\%$ in the NMO and $\sim 30\%$ in the IMO. The electron antineutrino survival probability only decreases in the NMO from $\sim 70\%$ due to diabaticity but increases from $\sim 0\%$ in the IMO. (2) At early times ($t\lesssim 0.6$ s), the line-of-sight dependence is present only for energies above a threshold (which varies between 1 and 5 MeV) and it is largely binary (all lines of sight can be grouped into those that traverse the deflagration plume and those that do not). For these early times, it is also clear the evolution of the survival probabilities appears to be smoothly energy dependent. This stands in contrast to the more chaotic variations seen at later times.  (3) At later times ($t\gtrsim 0.6$ s) and for neutrinos and NMO (Fig. \ref{fig:PeeEarthTblNH}) or for antineutrinos and IMO (Fig. \ref{fig:PBareeEarthTblIH}), the highly discontinuous density profile of the SN imprints itself upon the survival probabilities leading to variations of up to 40\% for neutrinos in the NMO and 70\% for antineutrino in the IMO. This variability occurs across the whole range of energies considered and persists for the full duration of the neutrino signal after $t>0.6$ s. (4) For late times ($t\gtrsim 0.6$ s) the turbulence and phase effects in the neutrinos if the mass ordering is inverted (Fig. \ref{fig:PeeEarthTblIH}) or in the antineutrinos if the ordering is normal (Fig. \ref{fig:PBareeEarthTblNH}) are much smaller in scale.


\subsection{The neutrino flux at Earth\label{sec:OscFlux}}
\begin{figure*}
\includegraphics[trim={0 1.8cm 0 1.2cm},clip,width=0.75\linewidth]{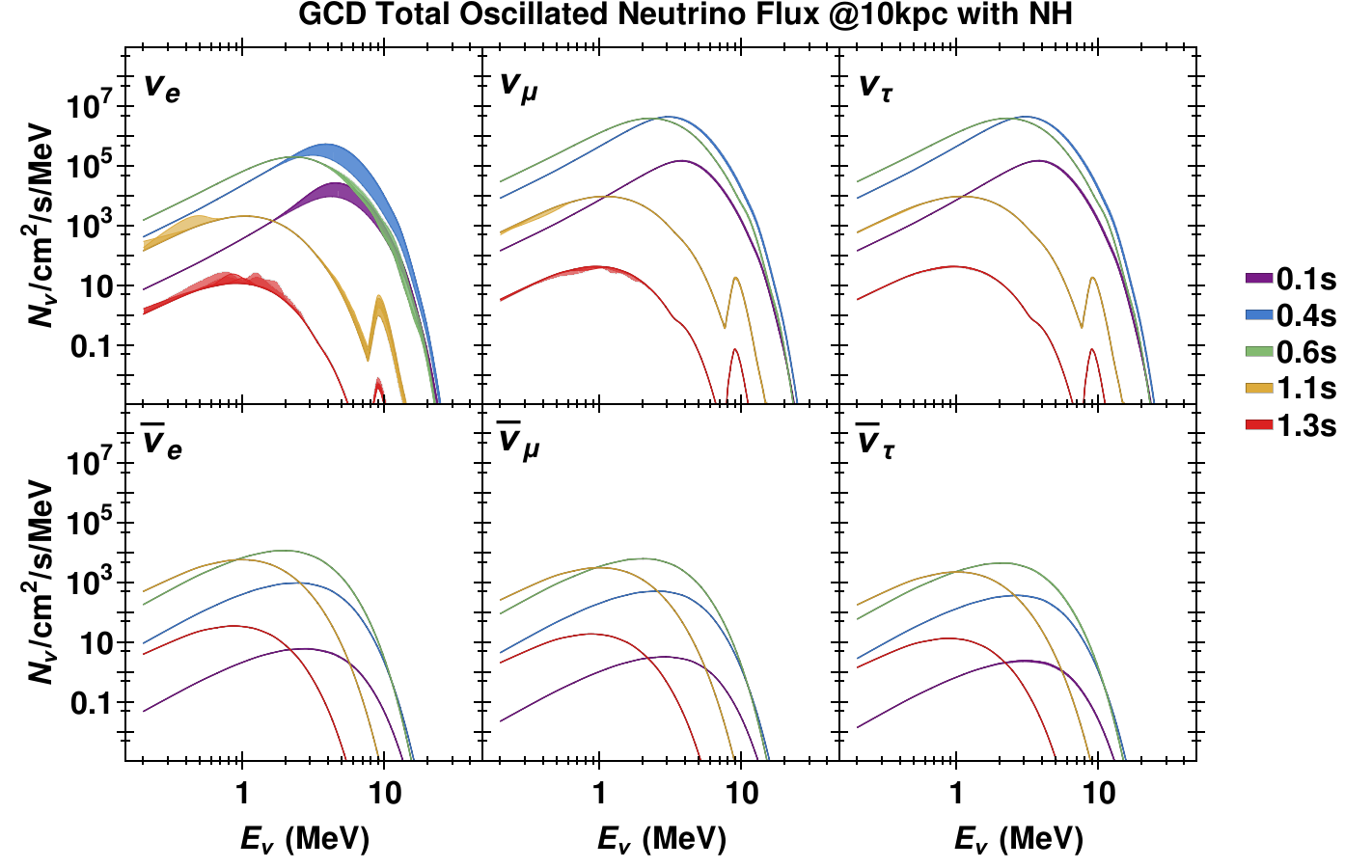}
\includegraphics[trim={0 0 0 1.2cm},clip,width=0.75\linewidth]{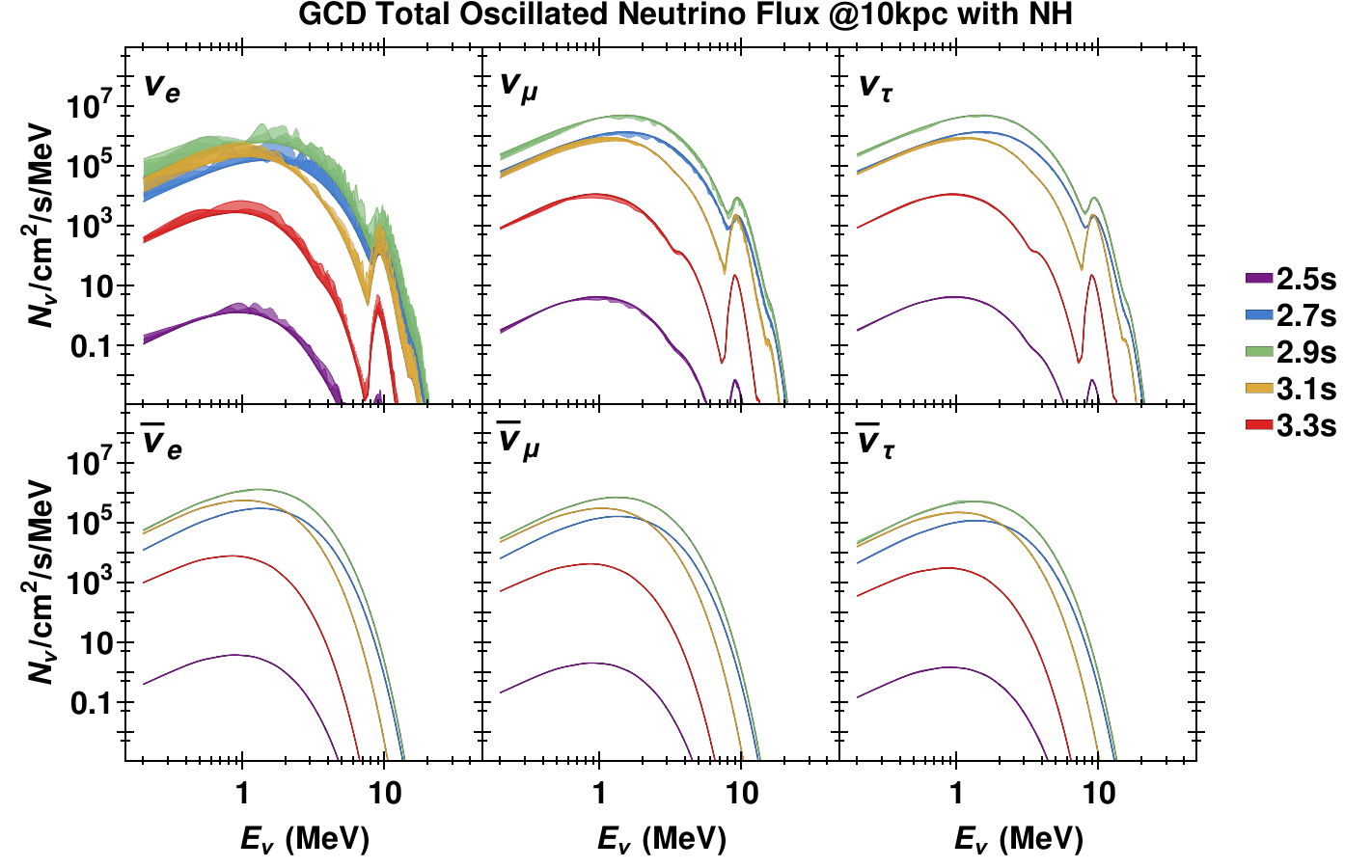}
\caption{The neutrino and antineutrino flux for each flavor from the GCD simulation at 10 kpc after oscillations for a NMO. Line thickness represents line-of-sight variation. In the top figure are the five snapshot times during the deflagration phase $t=0.1\;\text{s}$ to $t=1.3\;\text{s}$, the bottom figure shows the flux during the detonation phase $t=2.5\;\text{s}$ to $t=3.3\;\text{s}$. The color of each line indicates the snapshot time shown in the legend to the right of the figures.}
\label{fig:OscFluxNH}
\end{figure*}

\begin{figure*}
\includegraphics[trim={0 1.8cm 0 1.2cm},clip,width=0.75\linewidth]{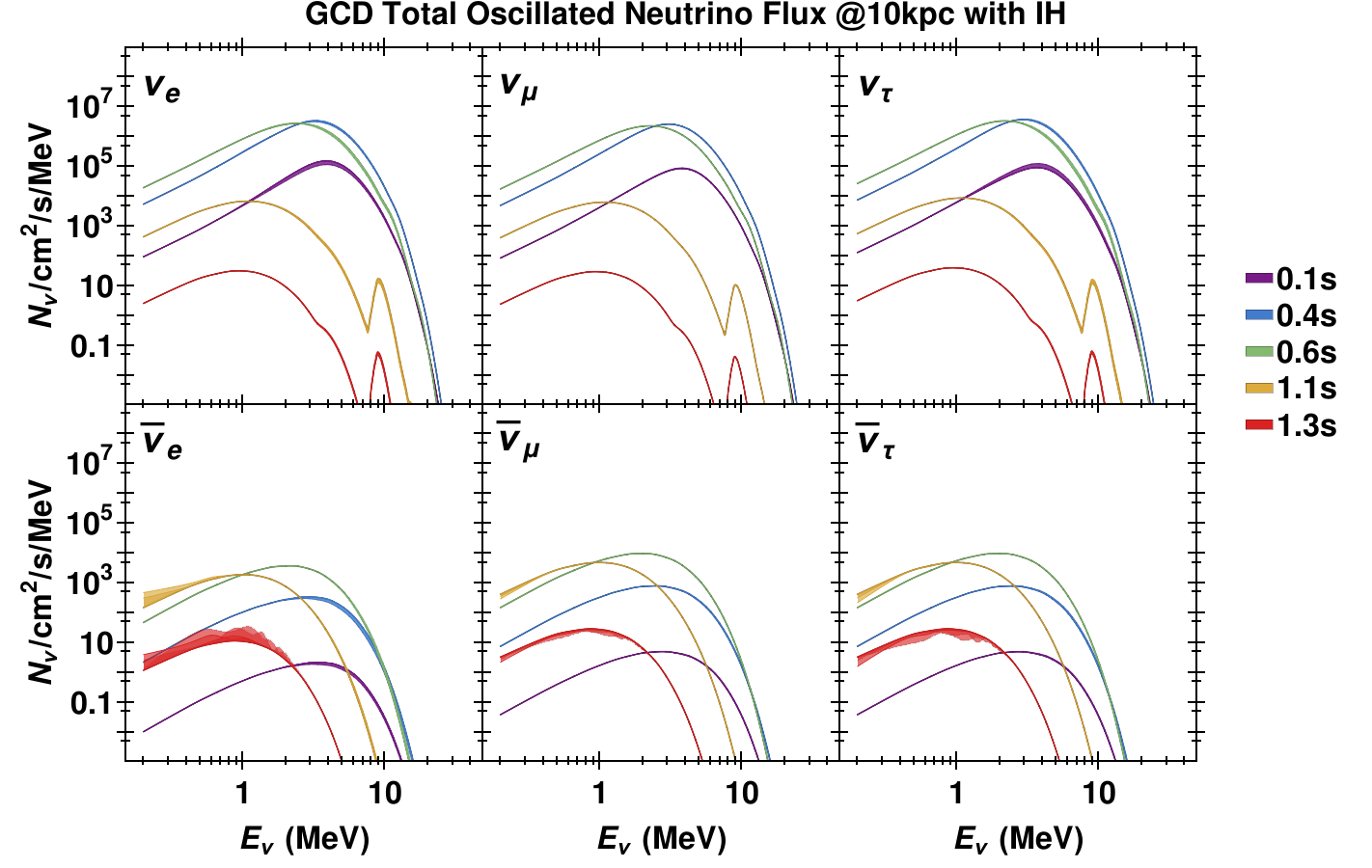}
\includegraphics[trim={0 0 0 1.2cm},clip,width=0.75\linewidth]{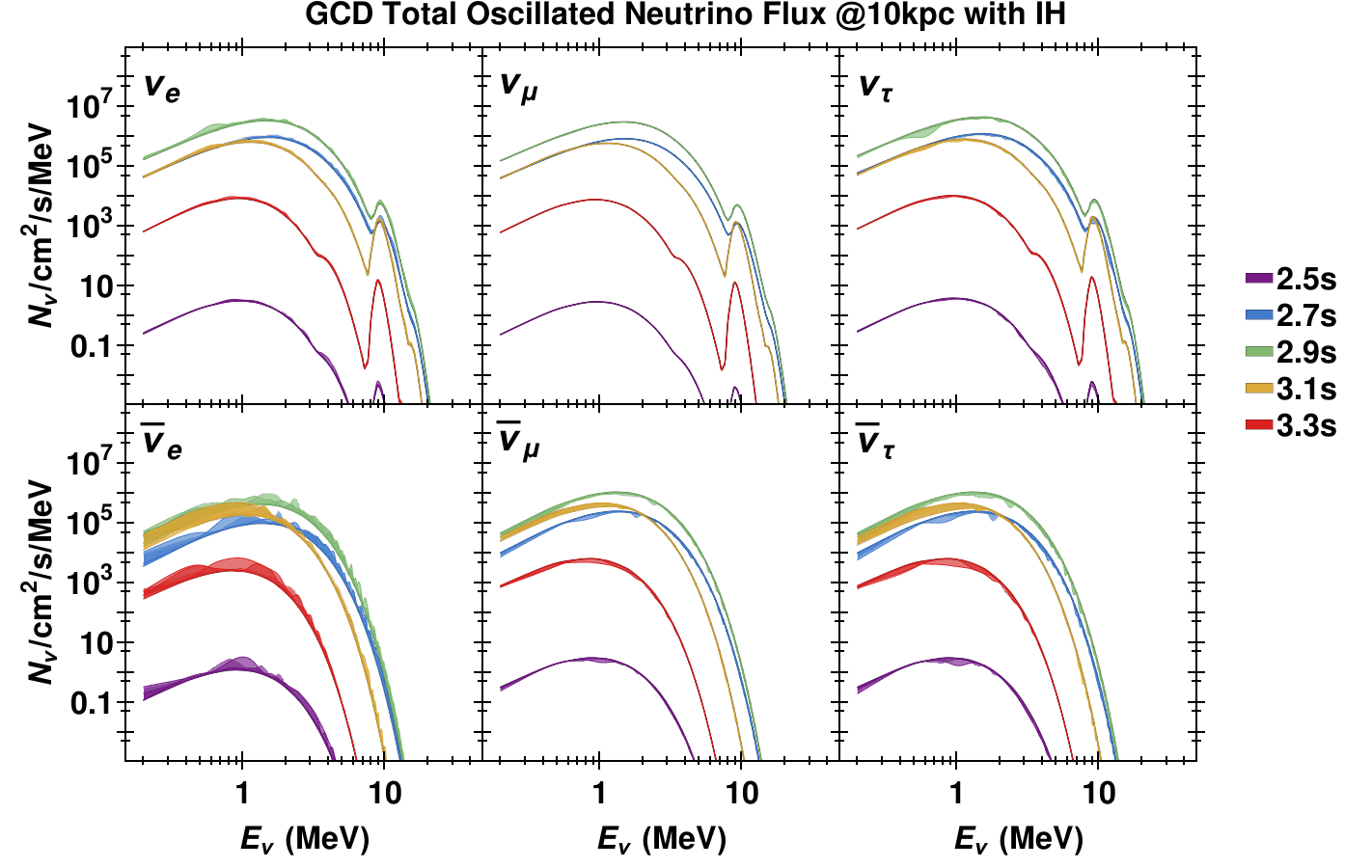}
\caption{The same as figure (\ref{fig:OscFluxNH}) but for an IMO.}
\label{fig:OscFluxIH}
\end{figure*}

\begin{figure*}
\includegraphics[trim={0 1.8cm 0 1.2cm},clip,width=0.75\linewidth]{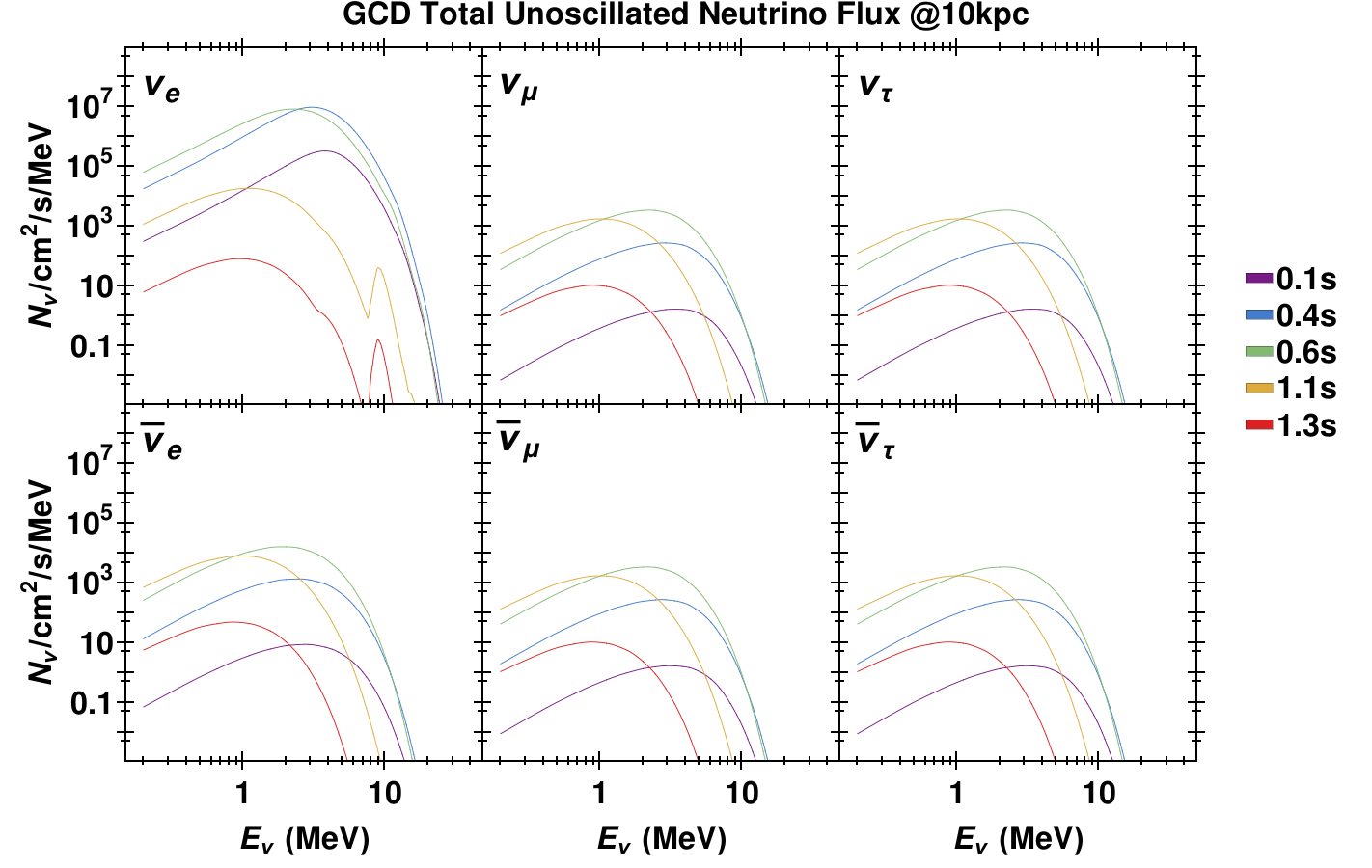}
\includegraphics[trim={0 0 0 1.2cm},clip,width=0.75\linewidth]{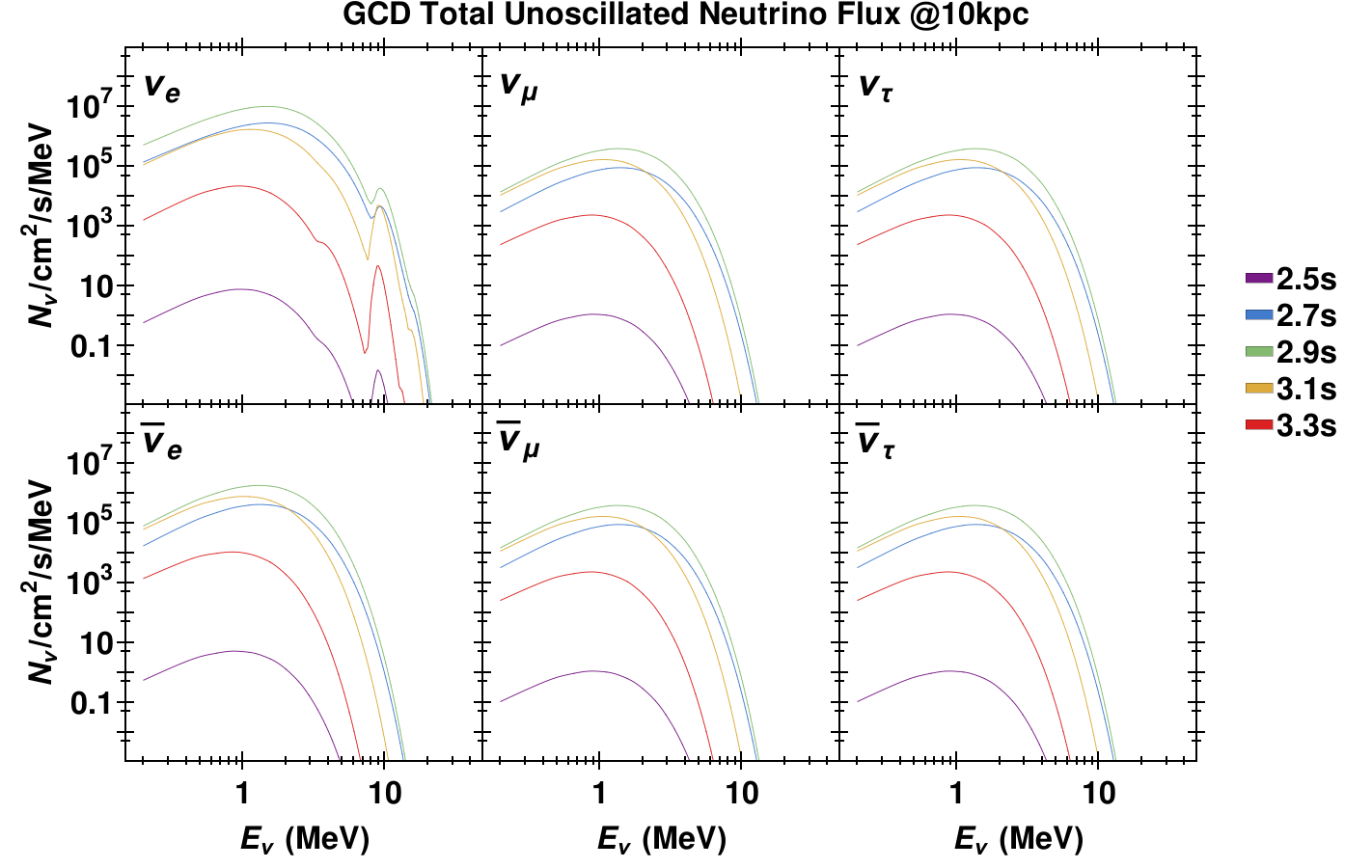}
\caption{The neutrino and antineutrino flux for each flavor from the GCD simulation at 10 kpc in the absence of flavor  oscillations. In the top figure are the five snapshot times during the deflagration phase $t=0.1\;\text{s}$ to $t=1.3\;\text{s}$, the bottom figure shows the flux during the detonation phase $t=2.5\;\text{s}$ to $t=3.3\;\text{s}$. 
The color of each line indicates the snapshot time shown in the legend to the right of the figures.}
\label{fig:UnoscFlux}
\end{figure*}

In order to calculate the flux seen by a detector, we combine the neutrino transformation probabilities at Earth (Sec. \ref{sec:PFlavor}) with the luminosity (Sec. \ref{sec:Production}) and then divide by a distance-dependent area factor. The distance used is $d =10\;{\rm kpc}$, the commonly used standard for Galactic supernovae and close to the most probable distance to a SN Ia of 9 kpc computed by Adams et al. \cite{2013ApJ...778..164A}. These fluxes at Earth are presented in Figs. \ref{fig:OscFluxNH} and \ref{fig:OscFluxIH} while the unoscillated neutrino flux at Earth is shown in Fig. \ref{fig:UnoscFlux} for comparison. In each case, the flux has been split into the two neutrino bursts; the deflagration burst is represented by the top six panels and the detonation burst is represented by the bottom six panels. For the oscillated figures, all ten of the angular neutrino trajectories considered are plotted and the space between the highest and lowest at a given energy is shaded in order to indicate the variation due to line-of-sight dependence.

For the deflagration neutrino burst (top plots of Figs. \ref{fig:OscFluxNH} to \ref{fig:UnoscFlux}) it is clear to see that the dominant $\nu_e$ flux has been largely oscillated into $\nu_\mu$ and $\nu_\tau$. The same is true for the detonation burst (bottom plots of Figs. \ref{fig:OscFluxNH} to \ref{fig:UnoscFlux}) though the effect is less severe because the $\nu_e$ flux is less dominant over the emitted $\nu_{\mu}$ and $\nu_{\tau}$ flux in the detonation burst epoch. In all cases, the 10 MeV production feature in the $\nu_e$ spectrum is seen in the $\nu_\mu$ and $\nu_\tau$ spectra due to the effect of the neutrino oscillations. It is worth pointing out that the 10 MeV feature is present in both the deflagration and the detonation burst. However, in the deflagration burst the 10 MeV peak develops midway through the burst. In contrast, for the detonation burst, the 10 MeV peak exists for the whole burst. Line-of-sight variation of the flux is almost an order of magnitude at some energies and the variation is most pronounced in the neutrino flux for NMO and in the antineutrino flux for IMO.

\section{Neutrino Detection \label{sec:NeutrinoDetection}}

\begin{figure*}[t]
\includegraphics[width=0.75\linewidth]{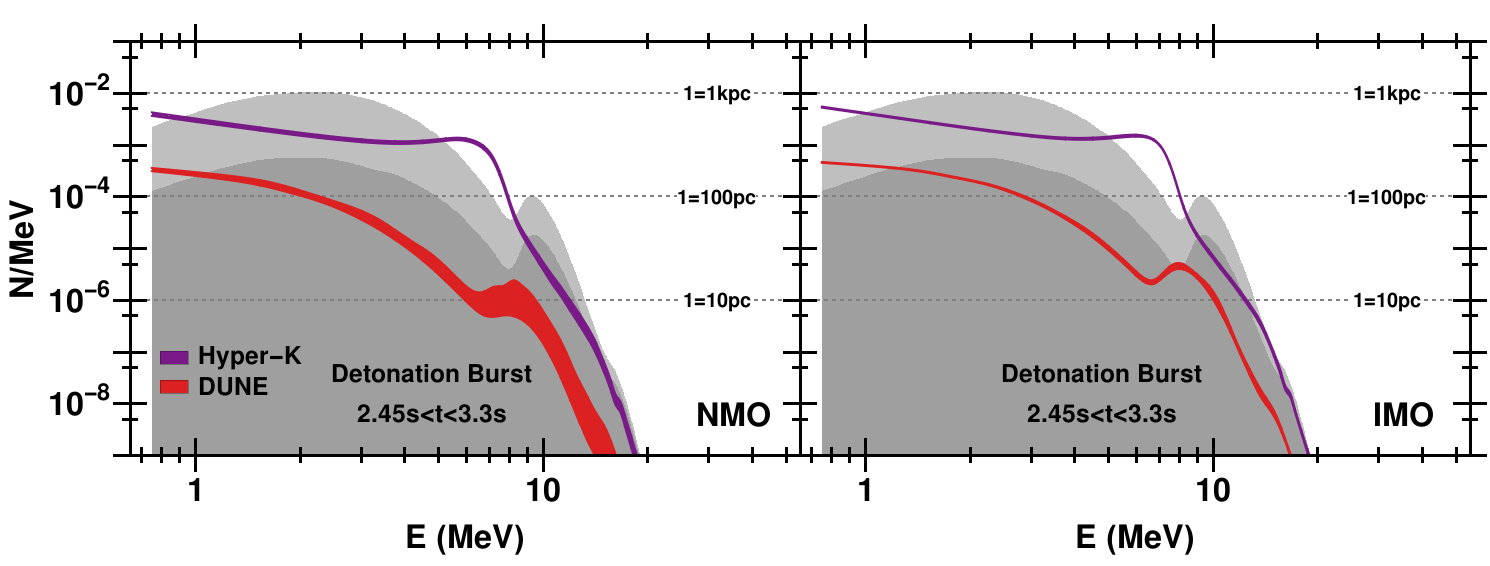}
\caption{The detector event spectrum in Hyper-K and DUNE for a GCD SN~Ia at 10 kpc integrated over the detonation burst i.e\ for $2.5\text{ s}<t<3.3\text{ s}$. The left column is for NMO and the right column is for IMO.
The thickness of the red and purple curves indicate the line-of-sight variation and the horizontal lines show how the rate (y axis) would shift for nearer SN. Detector energy smearing has been applied to the red and purple curves. The background gray curves represent the un-smeared event counts for comparison: the dark gray is for DUNE, the lighter gray for Hyper-K.}
\label{fig:EventsVsEnergy2DetectorsSmearing}
\end{figure*}

Finally, with the neutrino fluxes at Earth in hand, we can process them through a range of neutrino detectors and compute the expected event rates for the GCD SN~Ia in order to determine the detectability of the various production and oscillation features uncovered in the previous sections. The detectors under consideration are listed in Table \ref{table:Detectors} and are chosen because they are representative of various current and near-future neutrino detector technologies. The event rates in Super-K, Hyper-K\footnote{We are aware that the planned configuration of Hyper-K has changed. The old configuration is used here because the details of the new configuration are not yet available and for consistency with Paper I. To first order, the results presented here can be scaled by detector mass.}, DUNE, and JUNO were calculated using SNOwGLoBES \cite{SNOwGLoBES} and the event rate for IceCube was calculated according to the method described in Paper I. With the exception of Fig.~\ref{fig:EventsVsEnergy2DetectorsSmearing}, all of the results presented represent event counts where perfect detector efficiency is assumed and where energy smearing between the incoming neutrino and the detected lepton is ignored. The detectors in Table \ref{table:Detectors} are named with the suffix ``like'' to indicate that the configuration used only approximates the actual (or planned) detector. This is because, for the existing detectors, accurate configurations are not publicly available, and for future detectors, configuration information is not yet finalized. Although SNOwGLoBES gives results for each interaction channel separately, all the results we present are for the sum over all interaction channels. 

\renewcommand{\arraystretch}{1}
\begin{table}[ht!!!]\centering
		\begin{tabular}{l c c}
			\cline{1-3}
			Detector& Type& Mass (kt)\\ 
			\cline{1-3}
			Super-Kamiokande like: 30\%& Water Cherenkov&50\\ 
			\quad phototube coverage&  & \\  
			Hyper-Kamiokande like & Water Cherenkov& 560\\ 
			DUNE like detector&Liquid Ar&40\\ 
			JUNO like detector&Scintillator&20\\
   			IceCube & Water Cherenkov&3500$^*$\\            
			\cline{1-3}
		\end{tabular}
   \caption{Summary of the detectors under consideration.  Note that event rates simply scale by mass. $^*$For IceCube, the mass given is the effective mass used for the event rate calculation (see Appendix A of Paper I \cite{Wright2016}).}
	\label{table:Detectors}
\end{table}
\renewcommand{\arraystretch}{1}

\begin{figure*}[t]
\includegraphics[width=0.75\linewidth]{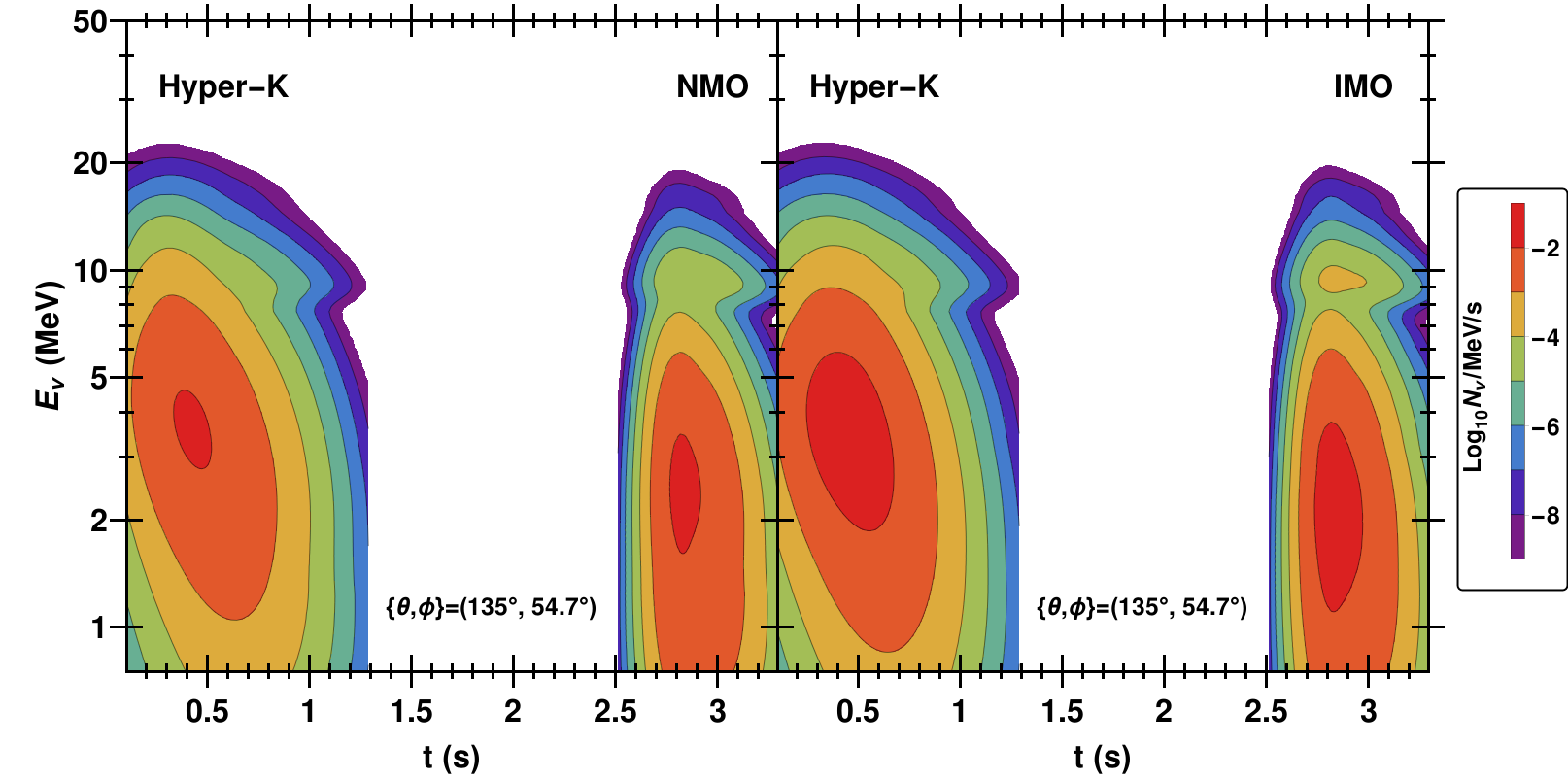}
\caption{A plot of the interaction event rate in Hyper-K for a particular line of sight for a GCD SN~Ia at 10 kpc. The left column is for NMO and the right column is for IMO.}
\label{fig:HyperKContours}
\end{figure*}

\begin{figure*}[t]
\includegraphics[width=0.75\linewidth]{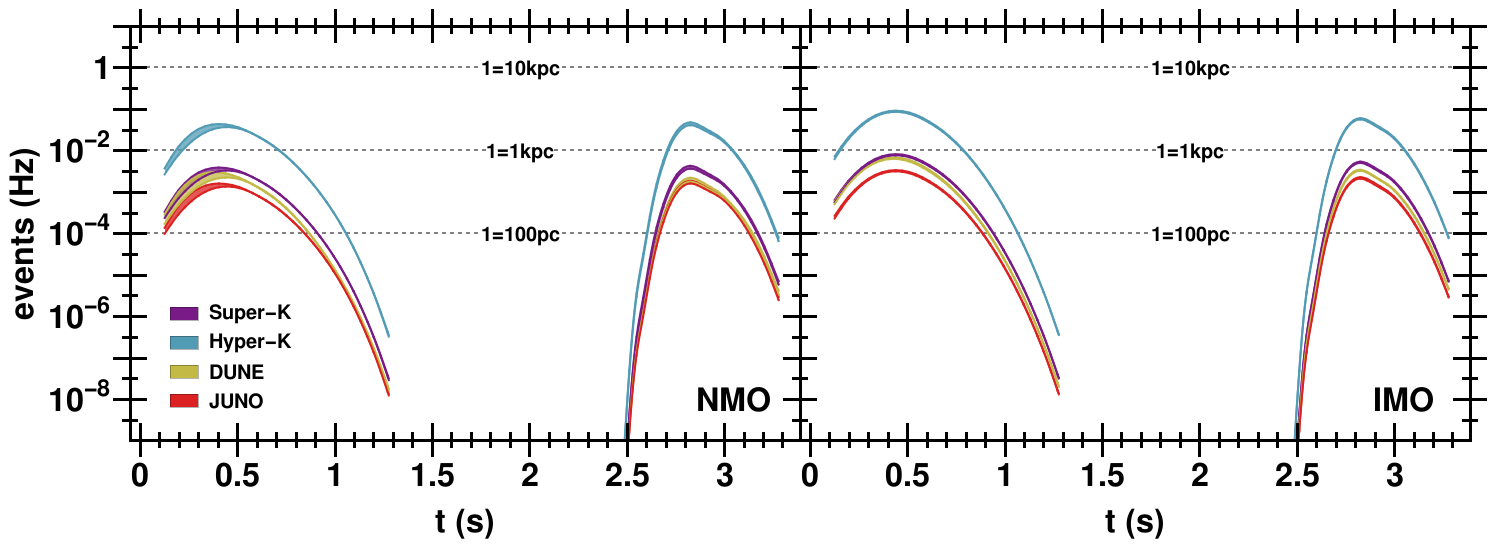}
\caption{The energy-integrated interaction event rate in Hyper-K, Super-K, DUNE, and JUNO for a GCD SN~Ia at 10 kpc. The curve thickness indicates line-of-sight variation and the horizontal lines show how the rate (y axis) would shift for nearer SN. The left column is for NMO and the right column is for IMO.}
\label{fig:EventsVsTime4Detectors}
\end{figure*}

Table \ref{table:Events} displays the number of interaction events in the detectors under consideration from a GCD SN~Ia at 10 kpc. In addition to all channels, these numbers are for all energy bins and integrated over the full duration of both neutrino bursts. Furthermore, the displayed event counts are the mean over all 10 lines of sight considered. The line-of-sight variation is $\sim5\%$ in NMO and $\sim2\%$ in IMO. 
The lack of significant line-of-sight dependence means there is little ambiguity and uncertainty in the interpretation of our results. It is remarkable that the total signal has such a small line-of-sight variation considering how asymmetric the GCD simulation is. This is a consequence of a lower line-of-sight variation during the time periods and at the energies of maximum neutrino emission. The table also shows the event rates in the same five detectors when we switch off neutrino flavor oscillations. Accounting for different detector masses and assumed distances, the number of events listed in this column are about a factor of 2-3 smaller than the number of events listed in Table 2 of Odrzywolek and Plewa for their Y12 model for comparable detectors.

The results in Table \ref{table:Events} demonstrate the well-known fact that neutrino flavor transformation generally reduce the signal expected from SNe~Ia without oscillations. For an IMO there are roughly only half as many events; for an NMO the number of events is reduced by a factor $\sim4$. The overall reduction due to oscillation is caused by transformation of the flavor which dominates the emission at the source, $\nu_e$, into other flavors at Earth. If the neutrinos propagated adiabatically then the electron neutrinos at the source would all but disappear into $\nu_{\mu}$ and $\nu_{\tau}$ for a NMO, and for an IMO we would observe only $30\%$ of the original $\nu_e$.
Even though we find the neutrinos do not propagate adiabatically at all energies and for all epochs, the diabaticity introduced by both the steep density gradients at the edge of the star and the multiple discontinuities in the deflagration ash is unable to increase the event rates by factors greater than a few. 

Clearly, a GCD SN~Ia at 10 kpc will not be visible in any of the detectors listed in the table. If we change the distance to the supernovae, the event rates scale as $(10~{\rm kpc}/d)^2$. Thus it would require the distance to the supernova be reduced to 1 kpc before Hyper-K and IceCube would only see a few interaction events. 
It is important to note that while the predicted number of interaction events in IceCube is similar to the number predicted in Hyper-K, the background separation is more demanding in IceCube. Hence, detection in IceCube would optimistically require a SN within ${\sim}50$ pc to be seen as a greater than $1\sigma$ deviation above the background.
For Super-K, JUNO, and DUNE we  would need a GCD SN~Ia at $\sim0.3$ kpc to see a few interaction events. According to Adams et al. \cite{2013ApJ...778..164A}, the probability of a SN~Ia being within 1 kpc is only ${\sim}0.2\%$. 

In addition to distance, we should not forget the event rate in a detector also scales by the detector mass (ignoring changes in detection efficiency). From the results in the table we see that one would need approximately a $10\;{\rm Mt}$ water Cherenkov detector to detect a galactic GCD SN~Ia at the most probable distance of 10 kpc. This is similar to the 5 Mt detector considered by Kistler {\it et al.} \cite{2011PhRvD..83l3008K}. 

\begin{figure*}[t]
\includegraphics[width=0.75\linewidth]{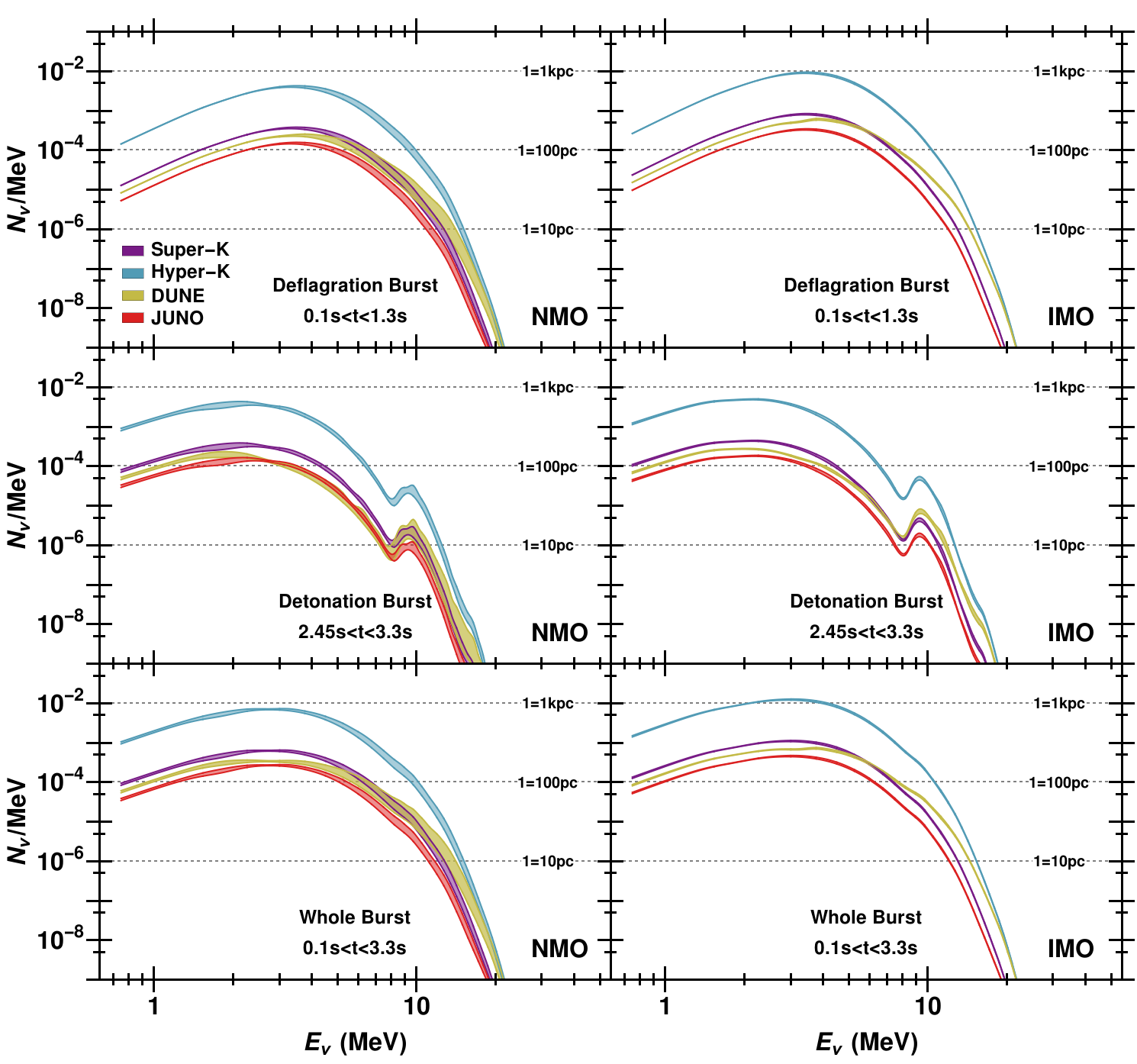}
\caption{The detector event spectrum in Hyper-K, Super-K, JUNO, and DUNE for a GCD SN~Ia at 10 kpc. The curve thickness and shading indicates line-of-sight variation and the horizontal lines show how the rate (y axis) would shift for nearer SN. The top row are for deflagration burst ($0.1\text{ s}<t<1.3\text{ s}$), the middle row for the detonation burst ($2.5\text{ s}<t<3.3\text{ s}$), and the bottom row are for the whole neutrino signal ($0.1\text{ s}<t<3.3\text{ s}$). The left column is for NMO and the right column is for IMO.}
\label{fig:EventsVsEnergy4Detectors}
\end{figure*}

\renewcommand{\arraystretch}{1}
\begin{table}\centering
		\begin{tabular}{l c c c}
			\cline{1-4}
			Detector&NMO&IMO&Unoscillated\\ 
			\cline{1-4}
			Super-Kamiokande&$0.0024 $&$0.0043 $&$0.0089$\\ 
			Hyper-Kamiokande&$0.0267 $&$0.0478 $&$0.0997$\\ 
			DUNE&$            0.0015 $&$0.0033 $&$0.0069$\\ 
			JUNO&$            0.0010 $&$0.0018 $&$0.0037$\\
			IceCube*&$        0.0208 $&$0.0325 $&$0.0689$\\
			\cline{1-4}  
		\end{tabular}
   \caption{Numbers of interactions per detector for each mass ordering and a SN at 10 kpc. These event counts are for the whole 3.3 s neutrino burst averaged over the ten lines-of-sight considered. The fourth column represents the number of interactions observed when neutrino oscillations are not taken into account.\newline * Note that the numbers of interactions quoted for IceCube are after background subtraction.}
	\label{table:Events}
\end{table}
\renewcommand{\arraystretch}{1}

Even though the neutrino detection probability of a galactic GCD SN~Ia is quite low, if the supernova is close and next-generation neutrino detectors are sufficiently large, it would be possible to go beyond total event rates and try to determine whether features in the signal can be discerned by examining the time and energy structure of the neutrino signal. Figure \ref{fig:HyperKContours} show the differential interaction event rate in Hyper-K as a function of time and energy. Both mass orderings are displayed and one particular line of sight was chosen (neither the line of sight along the ignition points of the deflagration nor detonation). This figure demonstrates a clear distinction between the deflagration burst ($0.1\text{ s}<t<1.3\text{ s}$) and the detonation burst ($2.5\text{ s}<t<3.3\text{ s}$). The peak energy during the deflagration burst is $\sim4$ and $\sim 2\;{\rm MeV}$ during the detonation burst. The 10 MeV production peak is mostly washed out during the deflagration peak but is clearly visible for the detonation peak. Lastly, the line-of-sight variation to Fig. \ref{fig:HyperKContours} is small and does not change the above observations. Similar figures for the other detectors we are considering show similar structure. 

We integrate the differential interaction event rates over energy and produce Fig. \ref{fig:EventsVsTime4Detectors} which shows the differential interaction event rate from a GCD SN~Ia at 10 kpc for Hyper-K, Super-K, DUNE, and JUNO. The line-of-sight variation is represented by the curve thickness, both mass orderings are shown, and the horizontal dotted lines indicate how the rate would shift for SNe at closer distances. This figure reveals that the line-of-sight variation is most prominent for $t<0.5$ s. At these early times, the line-of-sight variation is because the deflagration plume is very asymmetric and thus lines of sight that traverse the plume will result in markedly different signals compared to those lines of sight that do not traverse the deflagration plume. Once the deflagration engulfs the star, the variation decreases due to the overall increase in adiabaticity. Another interesting feature is that the detonation burst produces roughly as many events as the deflagration burst making this explosion-mechanism-differentiating feature just as visible as the signal as a whole. We also see the distinct separation of the deflagration and detonation bursts in the GCD scenario, with an estimated ${\sim}1.2$ s gap between the bursts. As pointed out by Odrzywolek and Plewa \cite{Odrzywolek2011a}, this double-peak time structure of the signal has great potential to distinguish the GCD explosion mechanism from other explosion mechanisms if the supernova occurs in sufficient proximity. 

Finally, we can integrate the differential interaction event rates over time and look at spectral features. 
The results are shown in Fig. \ref{fig:EventsVsEnergy4Detectors} and, as before, curve thickness represents line-of-sight variation, both mass orderings are displayed, and both bursts, as well as the spectrum obtained by combining them. 
As observed from Fig. \ref{fig:HyperKContours}, we see the peak energy for the deflagration burst is greater than that of the detonation burst. 
The line-of-sight variation is mostly confined to energies greater than 4 MeV for the deflagration burst. In contrast, the line-of-sight variation in the detonation burst is visible across the whole spectrum. Line-of-sight variation is also larger for NMO than for IMO.
Of particular note is the 10 MeV spectral bump discussed in Sec. \ref{sec:Production} which we find is only distinguishable in the detonation peak. However the figure indicates Super-K, JUNO, and DUNE would need a SN within 10 pc to detect the 10 MeV feature while Hyper-K would be able to distinguish the 10 MeV peak at $\sim 30$ pc. A supernova this near is quite unlikely. 

Thus far we have only reported interaction events where the detected energy is that of the incoming neutrino. A more realistic approach is to examine the events as a function of the energy of the detected final state interaction products, smeared according to detector resolution.  SNOwGLoBES provides event rates smeared according to distribution of final state products observed in a given detector.  Figure \ref{fig:EventsVsEnergy2DetectorsSmearing} displays how the results shown in Fig. \ref{fig:EventsVsEnergy4Detectors} will change with energy smearing taken into account. Only Hyper-K and DUNE are considered and the unsmeared interaction event energy is displayed as a gray background. The $x$ axis now represents the detected energy. The "peak" seen in the Hyper-K spectrum is not a displaced 10 MeV peak; rather it is a consequence of the approximate smearing function. However, the  10 MeV peak seen in the DUNE spectrum is the 10 MeV production feature that has been displaced to approximately 8 MeV because of the smearing. Therefore, our approximate smearing techniques show that energy smearing drastically reduces the visibility of the 10 MeV peak in Hyper-K but only shifts it for DUNE.

\section{GCD versus DDT \label{sec:Compare}}

In this last section, we turn our attention to comparing the neutrino signal from the DDT SN~Ia considered in Paper I to the GCD SN~Ia neutrino signal described above. The neutrino emission from both the DDT and GCD is comprised of an early burst from a period of deflagration burning which transitions to a second burst of neutrino emission from a detonation phase. The big difference between the two scenarios is the ratio of neutrino luminosity from the two phases and the time between them. This difference is shown in Table \ref{table:DDTvsGCDluminosities} where we list the peak neutrino luminosity as calculated for the two explosion mechanisms during the two phases and the times at which they occur. Firstly, the table shows the DDT model has more than an order of magnitude more luminosity when compared to the GCD model. This is reflected in both the maximum luminosity and the integrated luminosity. However we notice the peak luminosity during the detonation is much greater for the GCD model than for the DDT model. This means that, even though the DDT model emits more neutrinos, it is the GCD model that has the strongest detonation peak. Secondly, the time separation between the deflagration and the detonation burst is much larger for the GCD model than for the DDT model. After the overall event count difference, this time separation feature, if it can be detected, is a clear discriminating difference between the two models.

\renewcommand{\arraystretch}{1.5}
\begin{table}\centering
		\begin{tabular}{ c  c  c  c  c  c }
			\cline{1-6}
			\multirow{2}{*}{}&\multicolumn{2}{c}{Deflagration}&\multicolumn{2}{c}{Detonation}&Total\\ 
            \cline{2-6}
			Model&time&erg/s&time&erg/s&erg\\ 
			\cline{1-6}
			DDT&$0.53$ s&$5.1\times10^{49}$&$1.32$ s&$2.3\times10^{47}$&$2.0\times10^{49}$\\
			GCD&$0.45$ s&$2.3\times10^{48}$&$2.82$ s&$1.8\times10^{48}$&$1.2\times10^{48}$\\ 
			\cline{1-6}  
		\end{tabular}
   \caption{Comparison of the maximum luminosity between the DDT and the GCD SN~Ia simulations. The maximum luminosity is given for both the deflagration and detonation bursts. Also given is the total integrated luminosity for each model. It is important to note that in-between the deflagration and the detonation bursts, the GCD has very little neutrino emission but the DDT still has emission $\sim1.3\times10^{47}$ erg/s.}
	\label{table:DDTvsGCDluminosities}
\end{table}
\renewcommand{\arraystretch}{1} 

\begin{figure*}[t]
\includegraphics[trim={0 0 0 1.4cm},clip,width=0.75\linewidth]{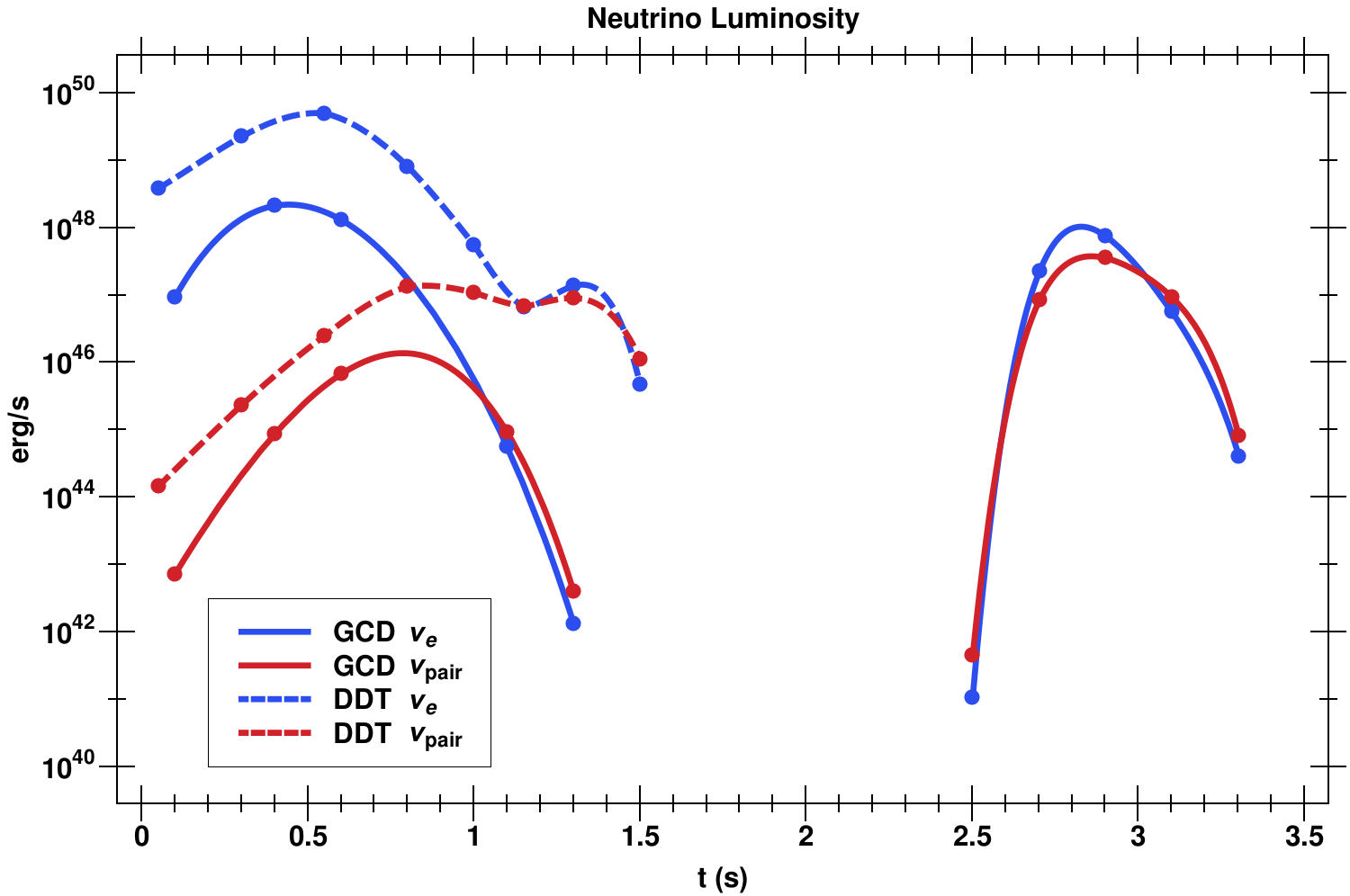}
\caption{A comparison of the neutrino luminosities for the DDT and GCD SN~Ia simulations. The layout and structure is similar to that of Fig. \ref{fig:TotalNeutrinoLuminosityVStime}}
\label{fig:DDTvsGCDluminosityVsTime}
\end{figure*}

The features mentioned above are also clear from Fig. \ref{fig:DDTvsGCDluminosityVsTime} where the neutrino luminosity is plotted against time. 
In this figure, the emission arising from weak and pair annihilation processes are shown separately. It is again clear that the DDT model has a greater luminosity than the GCD model and that the DDT model has a shorter gap between the deflagration and detonation bursts. Figure \ref{fig:DDTvsGCDluminosityVsTime} shows another difference, namely that the DDT model still has significant neutrino emission in between the two bursts while the GCD model has practically no emission in between its deflagration and detonation bursts. An important feature that both the DDT and the GCD models share is that the deflagration burst is dominated by weak emission; however during the detonation burst, weak and thermal emission are equally important in both models.


\begin{figure*}[t]
\includegraphics[trim={0 0 0 1.2cm},clip,width=0.75\linewidth]{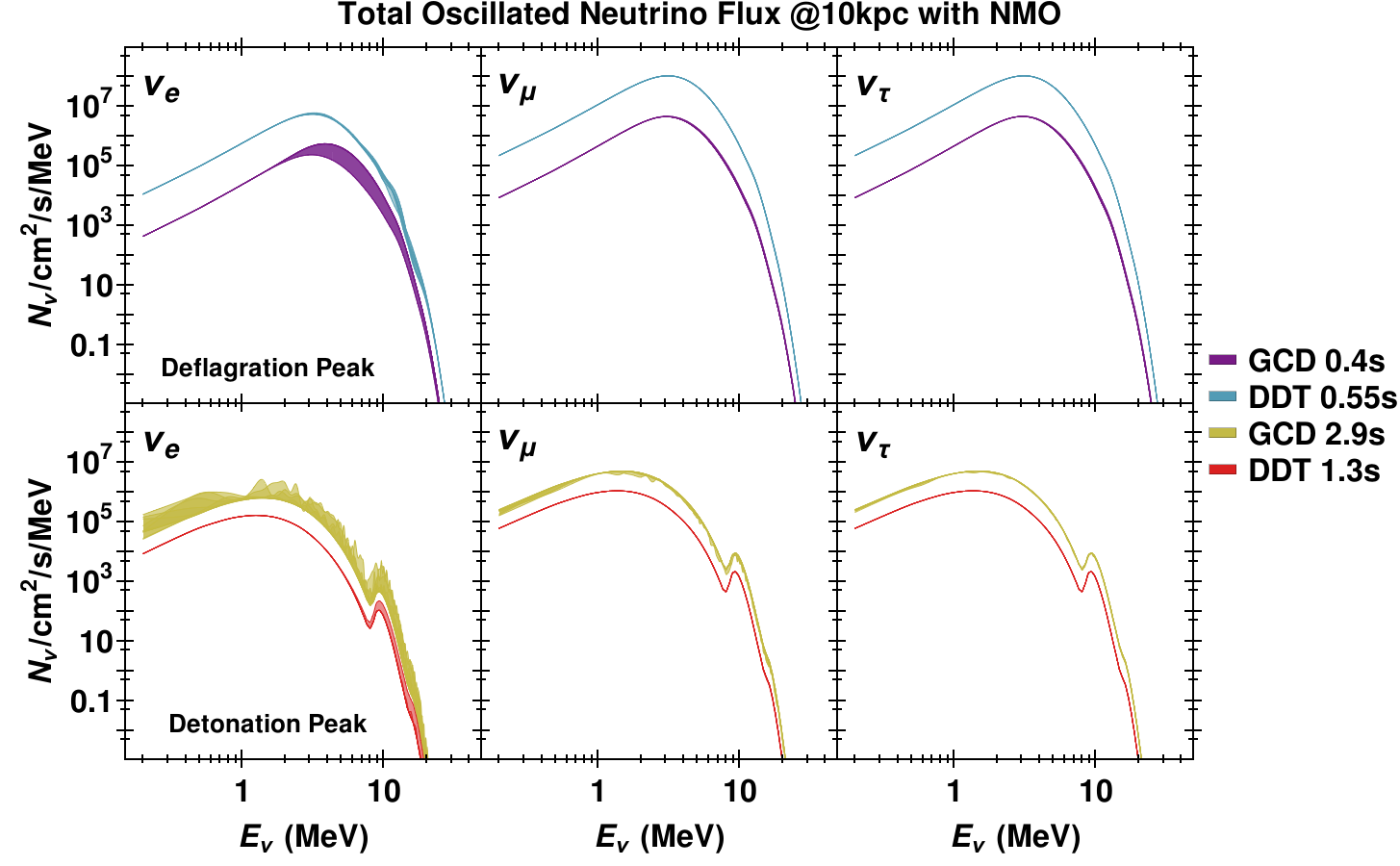}
\caption{A comparison of the differential neutrino flux on Earth for the DDT and GCD SN~Ia simulations. The fluxes include the effects of oscillations for a NMO and $d = 10$ kpc. The three columns are for the three flavors of neutrinos. The top row shows the spectra at two snapshots during the deflagration burst of each simulation ($t_\text{GCD}=0.4$ s and $t_\text{DDT}=0.55$ s) and the bottom row shows the spectra at two snapshots during the detonation burst of each simulation ($t_\text{GCD}=2.9$ s and $t_\text{DDT}=1.3$ s). The simulation and the snapshot times are shown in the legend to the right. For each spectrum shown, the thickness of the curve indicates the range due to the line-of-sight variation through each calculation.}
\label{fig:DDTvsGCDspectrum}
\end{figure*}

\begin{figure*}[t]
\includegraphics[width=0.75\linewidth]{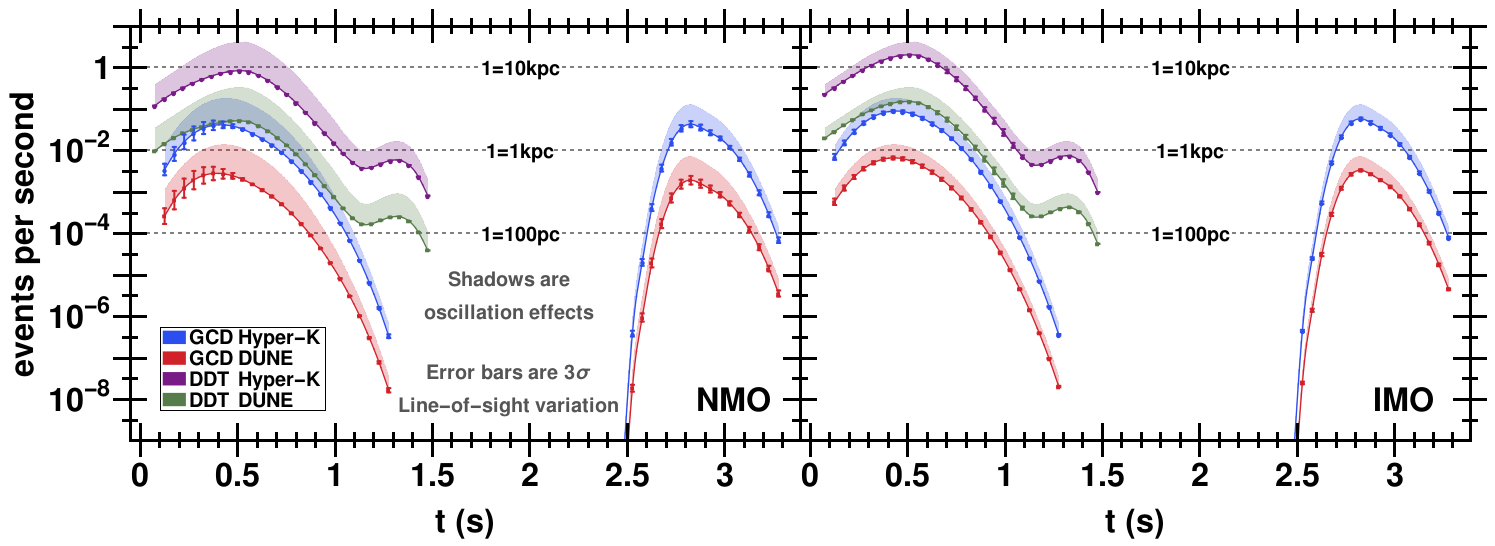}
\caption{The time evolution of the event rate in Hyper-K and DUNE for DDT and GCD simulations of a SN Ia at 10kpc. The shadowed regions represent the effects of neutrino oscillation such that the top of each shadowed region represents the associated unoscillated event rate. The lines represent the mean event rate across all neutrino trajectories considered. Similarly, the error bars represent the $3\sigma$ deviation due to line-of-sight variations in the event rates. The left (right) plot is for normal (inverted) mass ordering. The horizontal lines show how the event rate would change if the supernova occurred at a closer distance.}
\label{fig:DDTvsGCDeventsVsTime}
\end{figure*}

But the neutrino emission is not the only way the two models differ. 
The degree to which neutrino oscillations affect the flux from both the DDT and the GCD models is shown Fig. \ref{fig:DDTvsGCDspectrum}. This figure only displays the differential spectrum of $\nu_e$, $\nu_\mu$, and $\nu_\tau$ at the peaks of the deflagration and detonation bursts for each of the models. The curve thickness denotes line-of-sight variation, the mass ordering is normal, and the distance to the SN is set to 10 kpc. It is clear that the DDT has the most flux during the deflagration phase and the GCD has the most flux during detonation. Also clear is the absence of the 10 MeV peak from the deflagration peak for both models, while the 10 MeV peak is present in the detonation peak of both models. The most important feature of Fig. \ref{fig:DDTvsGCDspectrum} is found in comparing the line-of-sight variation of the two models. The DDT model has very little line-of-sight variation during the peak emission. Conversely, the GCD has much more line-of-sight variation during its peak emissions. This is not surprising when one considered how much more the GCD model departs from spherical symmetry when compared to the DDT model.

The differences between the two models in both the emitted spectra and flavor oscillations are preserved in the neutrino spectra at Earth and the event rates one expects in the various detectors we have considered. 
Figure \ref{fig:DDTvsGCDeventsVsTime} shows the event rate in Hyper-K and DUNE for both the DDT and the GCD models. Both mass orderings are presented, the error bars denote $3\sigma$ line-of-sight variation, the SN is at 10 kpc, and the horizontal lines indicate how the rates would change for nearer SN. 
This figure again indicates the line-of-sight variation is 10 times greater in GCD than DDT but still only a <10\% correction. The line-of-sight variation in the GCD rates are most pronounced during $t<0.5$ s while the DDT model has not appreciable variation at these times. Oscillation effects are roughly equally important in both models and the overall trend is that neutrino oscillations reduce the expected signal. The shaded area represents the effect of oscillations, i.e. the top of the shaded area corresponds to the rate if oscillations are ignored. It is striking that the main features observed from Table \ref{table:DDTvsGCDluminosities} and Fig. \ref{fig:DDTvsGCDluminosityVsTime} are clearly visible in Fig. \ref{fig:DDTvsGCDeventsVsTime}. Namely, that the DDT has greater event rates and the GCD has a greater separation between bursts. 

\begin{figure*}
\includegraphics[width=0.75\linewidth]{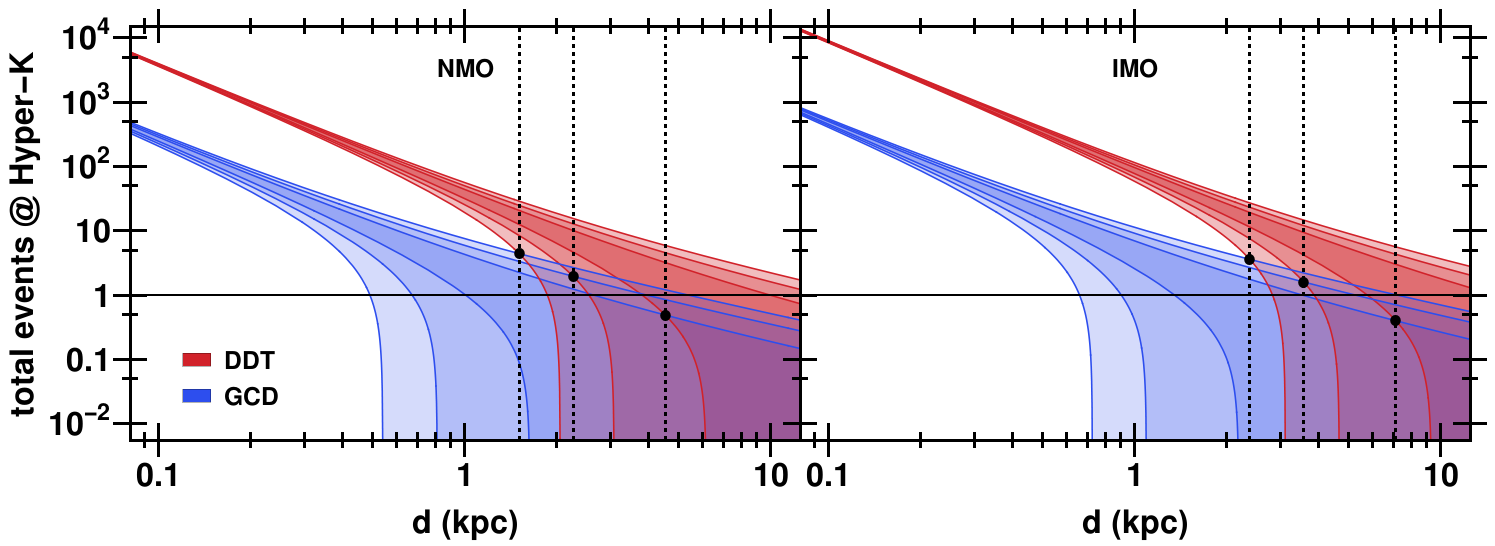}
\caption{The total number of events predicted to be measured in the Hyper-K detector as a function of the distance to the SN for both the DDT (red) and the GCD (blue) models. The left panel is for NMO and the right panel is the IMO. The shaded regions represent the event count range given a $1\sigma$ to $3\sigma$ Poisson error (also including a small error associated with line-of-sight uncertainty). The smallest, darkest shaded area represents the $1\sigma$ event range and the largest, lightest shaded area represents the $3\sigma$ event range for each model. The black dots show the distance at which the DDT total event count can be distinguished from the GCD total event count for a given $\sigma$ confidence. For NMO these distances are, from left to right, $\left(d_{3\sigma},d_{2\sigma},d_{1\sigma}\right) = \left(1.51, 2.26, 4.52\right)$ kpc. For IMO, these distances are, from left to right, $\left(d_{3\sigma},d_{2\sigma},d_{1\sigma}\right) = \left(2.38, 3.57, 7.13\right)$ kpc.}
\label{fig:DDTvsGCDeventsVsDistance}
\end{figure*}

\begin{figure*}[t]
\includegraphics[width=0.75\linewidth]{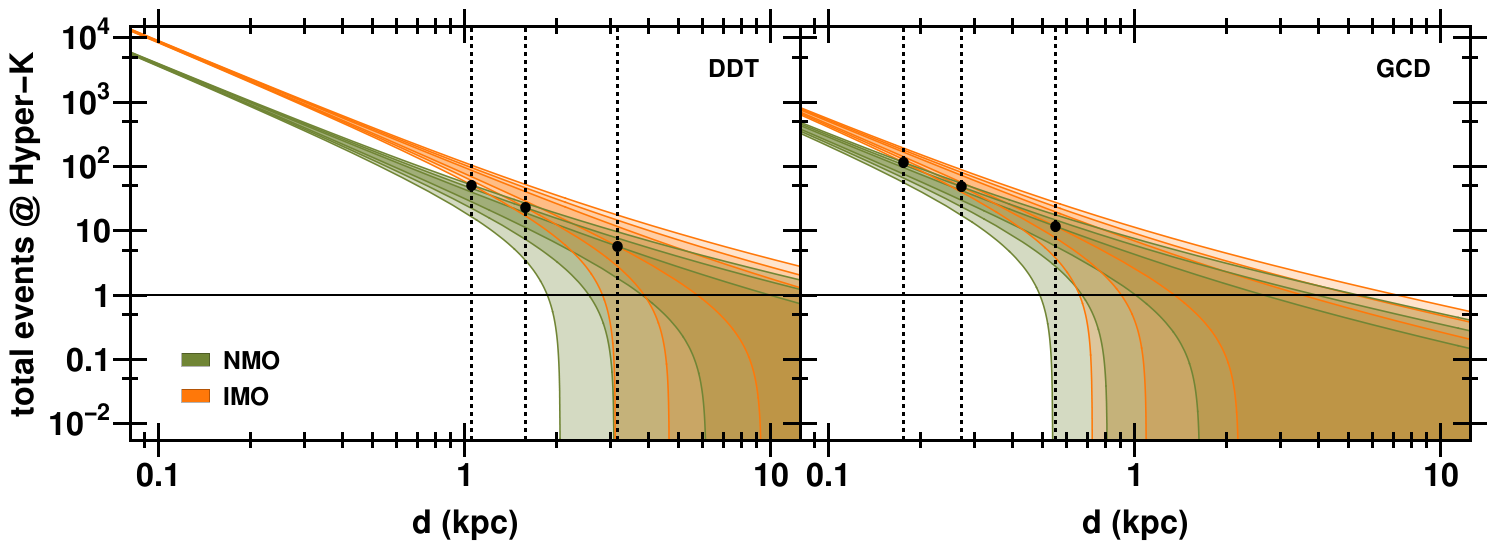}
\caption{The total number of events predicted to be measured in the Hyper-K detector as a function of the distance to the SN for both NMO (green) and IMO (orange). The left panel is for the DDT model and the right panel is the GCD model. The shaded regions represent the event count range given a $1\sigma$ to $3\sigma$ Poisson error (also including a small error associated with line-of-sight uncertainty). The smallest, darkest shaded area represents the $1\sigma$ event range and the largest, lightest shaded area represents the $3\sigma$ event range for each model. The black dots show the distance at which NMO can be distinguished from IMO for a given $\sigma$ confidence. For DDT these distances are, from left to right, $\left(d_{3\sigma},d_{2\sigma},d_{1\sigma}\right) = \left(1.05, 1.58, 3.17\right)$ kpc. For GCD these distances are, from left to right, $\left(d_{3\sigma},d_{2\sigma},d_{1\sigma}\right) = \left(0.18, 0.27, 0.55\right)$ kpc.}
\label{fig:NMOvsIMOEventsVSDistance}
\end{figure*}

The overall event rate difference between the two models can be used to discriminate between them. In Fig. \ref{fig:DDTvsGCDeventsVsDistance}, we show the 
event counts in Hyper-K for the DDT and GCD models as a function of distance plus $1\sigma$ to $3\sigma$ uncertainty from Poisson noise and line-of-sight uncertainty. Assuming the distance to the supernova is known, the event rate alone can discriminate between the models at $2\sigma$ if the distance to the supernovae is less than $d \leq 2.26 \;{\rm kpc}$ for NMO and less than $d \leq 3.57 \;{\rm kpc}$ for IMO. 

Similarly to Fig. \ref{fig:DDTvsGCDeventsVsDistance}, in Fig. \ref{fig:NMOvsIMOEventsVSDistance} the overall event rate difference between the two neutrino mass orderings can be used to discriminate between them. In Fig. \ref{fig:NMOvsIMOEventsVSDistance}, we show the event counts in Hyper-K for the two neutrino mass hierarchies as a function of distance plus $1\sigma$ to $3\sigma$ uncertainty from Poisson noise and line-of-sight uncertainty. Assuming the distance to the supernova is known, the event rate alone can discriminate between NMO and IMO at $1\sigma$ if the distance to the supernovae is less than $d \leq 3.17 \;{\rm kpc}$ for DDT and less than $d \leq 0.55 \;{\rm kpc}$ for GCD.

\begin{figure*}
\includegraphics[width=0.75\linewidth]{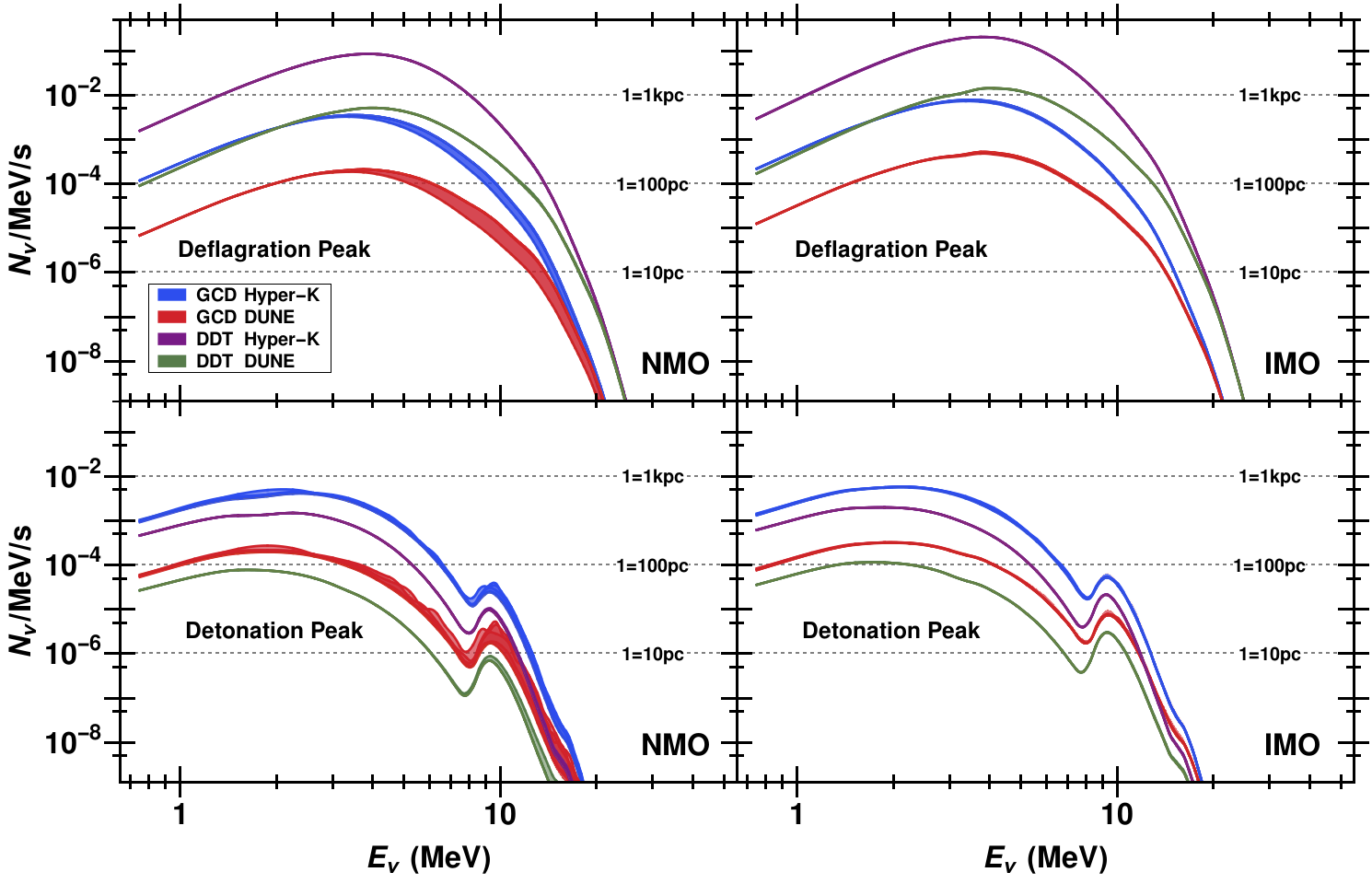}
\caption{The time evolution of the event rate spectra in Hyper-K and DUNE for DDT and GCD simulations of a SN Ia at 10kpc. The lines thickness represents the effect of line-of-sight variation. The left (right) plots are for normal (inverted) mass ordering. The horizontal lines show how the event rate would change if the supernova occurred at a closer distance. The top row represents the rate averaged over the deflagration burst and the bottom row represents the rate averaged over the detonation burst.}
\label{fig:DDTvsGCDeventsVsEnergy}
\end{figure*}

Lastly, Fig. \ref{fig:DDTvsGCDeventsVsEnergy} compares the event spectrum at both the deflagration and the detonation peaks for both models. This differential event rate shows the same features discussed in relation to Fig. \ref{fig:DDTvsGCDspectrum}. The new feature here is that we can quantify how much line-of-sight variation remains in the final interaction event rate spectrum. We see that even though there seems to be a significant amount of variation at higher energies, the variation near the peak of the emission is lower. Consequently, the line-of-sight variation for the GCD model is just a few percent and for the DDT model only a few tenths of a percent when the total rate is considered. Thus, even though the GCD model has an order of magnitude more line-of-sight variation, the overall effect on the detected signal is small. Figure \ref{fig:DDTvsGCDeventsVsEnergy} shows that, apart from overall normalization, the neutrino spectra are very similar between the DDT and the GCD models. This is remarkable considering how very different the two explosions mechanisms are.

\section{Conclusion \label{sec:Conclusion}}

In this paper, we have studied neutrinos from a gravitationally confined detonation scenario for a SN Ia in order to determine the features which might allow us to discriminate between this and other explosion mechanisms. \WW{Compared to earlier studies e.g.\~ by Odrzywolek \& Plewa \cite{Odrzywolek2011a}, we have added a number of refinements to the calculation. We adopt a 3D SNe~Ia simulation where previously 2D simulation was used, we calculate neutrino emission including more detailed spectral information, we include the effects of neutrino oscillations through the supernova by numerically solving the three-flavor evolution equations as a function of time, energy and for ten lines of sight, and we use a more complete detector modeling code \textsc{SNOwGLoBES} to compute the expected signal as a function of time and energy in a selection of detectors and for a variety of distances to the SN.}

The results show that the GCD SN~Ia produces two neutrino bursts. The first is associated with deflagration burning with a peak luminosity of $2.3\times10^{48}$ erg/s at $t=0.45$ s. The second burst is associated with detonation burning with a peak luminosity of $1.2\times10^{48}$ erg/s at $t=2.82$ s. There is very little neutrino emission in between the two bursts. We also find a 10 MeV $\nu_e$ spectral feature associated with electron capture on copper appears at $\sim 1$ s during the deflagration burst which persists for the entirety of the detonation burst. 

Neutrino oscillations introduce significant flavor conversion and are very line-of-sight dependent due to the discontinuity-ridden density profile and the asymmetrical explosion. The oscillations also deviate from adiabatic evolution across much of the time and energy parameter space. However, even though the oscillations show large line-of-sight dependence, the effect of line-of-sight variation on the total number of detected events is only a few percent. 

The calculated interaction event rates show that, for a GCD SN Ia at 1 kpc, Hyper-K and IceCube would only see a few interaction events. While Super-K, JUNO, and DUNE would need a GCD SN Ia at ${\sim}0.3$ kpc to see a few interaction events. Thus the conclusion is that one would either need an improbably close SN or order-of-magnitude larger detectors than current and near future proposals to confidently observe the neutrinos from a SN Ia with a GCD explosion mechanism. 

A comparison of the neutrino signals from the DDT SN and the GCD SN reveals several similar features between them. The line-of-sight analysis shows that neither model would suffer from a great amount ($<10\%$) of line-of-sight variation of the total detected events even though the GCD explosion is much more aspherical than the DDT. In both cases, inverted mass ordering yields more events than normal ordering but both are less then the case of no neutrino oscillations. The 10 MeV neutrino spectral feature that is found in both DDT and GCD is difficult to detect even for next generation neutrino detectors because it appears at late times when the luminosity is low. However the comparison between the DDT and GCD also reveals two very important, explosion model distinguishing, features. Firstly, the DDT SN has an order of magnitude more overall neutrino emission than the GCD SN. If the distance to the supernova can be determined then the large difference in event rates - assuming events are detected - will allow discrimination between the two models. If the supernova is sufficiently close that a star which explodes as a GCD SN Ia produces events in a detector, the distinctive separation in time between the deflagration and detonation burst of this scenario will also allow discrimination between the two scenarios independent of the overall event rate diagnostic. 
And finally, SN Ia can also provide information about neutrino properties if the explosion mechanism and distance are known. We find, if the mechanism and distance are known, the overall event rate can be used to determine the neutrino mass ordering.

\section*{Acknowledgments}
We are grateful to Evan O'Connor for his help with the implementation of \textsc{NuLib}.
This work was supported at NC State by Department of Energy Grants No. DE-SC0006417 and No. DE-FG02-10ER41577. Seitenzahl acknowledges support from the Australian Research Council Laureate Grant No. FL0992131. Scholberg's research activities are supported by the U.S. Department of Energy and the National Science Foundation. Ohlmann acknowledges support from Studienstiftung des Deutschen Volkes. The work of Ohlmann and R{\"o}pke is supported by the Klaus Tschira Foundation. Ohlmann, R{\"o}pke, and Seitenzahl thank the DAAD/Go8 German-Australian exchange programme (DAAD project 57135425) for travel support. 


\bibliography{main}

\end{document}